\documentclass[submission,copyright,creativecommons]{eptcs}
\providecommand{\event}{SYNT 2017} 

\usepackage{underscore}
\usepackage{graphicx}
\usepackage{adjustbox}
\usepackage{array}
\usepackage{booktabs}
\usepackage{caption}    
\usepackage{subcaption} 
\usepackage{multirow}
\usepackage{tikz}

\title{SyGuS-Comp 2017: Results and Analysis}
\author{
Rajeev Alur 
\institute{University of Pennsylvania}
\and
Dana Fisman
\institute{Ben-Gurion University}
\and
Rishabh Singh
\institute{Microsoft Research, Redmond}
\and
Armando Solar-Lezama
\institute{Massachusetts Institute of Technology}
}
\def\titlerunning{SyGuS-Comp 2016: Results and Analysis}
\def\authorrunning{R. Alur, D. Fisman, R. Singh \& A. Solar-Lezama}
\def\copyrightholders{Alur, Fisman, Singh \& Solar-Lezama}
\begin{document}
\maketitle

\newcommand{\commentout}[1]{}
\newcommand{\alc}{\textsc{Alchemist-cs}}
\newcommand{\alccsdt}{\textsc{Alchemist-csdt}}
\newcommand{\cvc}{\textsc{CVC4}}
\newcommand{\cvclast}{\ensuremath{\textsc{CVC4}_{2016}}}
\newcommand{\enum}{\textsc{Enumerative}}
\newcommand{\skac}{\textsc{Sketch-ac}}
\newcommand{\ice}{\textsc{Ice-dt}}
\newcommand{\toast}{\textsc{SosyToast}}
\newcommand{\stoch}{\textsc{Stochastic}}
\newcommand{\cvcnew}{\ensuremath{\textsc{CVC4}_{2017}}}
\newcommand{\eusolver}{\textsc{EUSolver}}
\newcommand{\eusolverlast}{\ensuremath{\textsc{EUSolver}_{2016}}}
\newcommand{\eusolvernew}{\ensuremath{\textsc{EUSolver}_{2017}}}
\newcommand{\euphony}{\textsc{Euphony}}
\newcommand{\ethree}{\textsc{e3solver}}
\newcommand{\dryd}{\textsc{DryadSynth}}
\newcommand{\lig}{\textsc{LoopInvGen}}
\newcommand{\sygus}{SyGuS}
\newcommand{\comp}{SyGuS-Comp}

\newcolumntype{R}[2]{%
	>{\adjustbox{angle=#1,lap=\width-(#2)}\bgroup}%
	l%
	<{\egroup}%
}
\newcommand*\rot{\multicolumn{1}{R{90}{1em}}}


\begin{abstract}
\emph{Syntax-Guided Synthesis (SyGuS)} is
the computational problem of finding an implementation $f$ that
meets both a semantic constraint
given by a logical formula $\varphi$ in a background theory $T$,
and a syntactic constraint given by a grammar $G$, which specifies the allowed set of
candidate implementations.
Such a synthesis problem can be formally defined in SyGuS-IF,
a language that is built on top of SMT-LIB.

The \emph{Syntax-Guided Synthesis Competition (\comp)} is an
effort to facilitate, bring together and accelerate research and development of efficient
solvers for SyGuS by providing a platform for evaluating different synthesis
techniques on a comprehensive set of benchmarks. 
In this year's competition six new solvers competed on over 1500 benchmarks.
This paper presents and analyses the results of \comp'17.
\end{abstract}

\section{Introduction}
\label{sec:intro}

The \emph{Syntax-Guided Synthesis Competition} (SyGuS-Comp) is an annual competition aimed to provide an objective platform for comparing different approaches for solving the \emph{Syntax-Guided Synthesis} (SyGuS) problem. A SyGuS problem takes as input a logical specification $\varphi$ for what a synthesized function $f$ should compute, and a grammar $G$ providing syntactic restrictions on the function $f$ to be synthesized. Formally, a solution to a SyGuS instance $(\varphi,G,f)$ is a function $f_{imp}$ that is expressible in the grammar $G$ such that the formula $\varphi[f/f_{imp}]$ obtained by replacing  $f$ by $f_{imp}$ in the logical specification $\varphi$ is valid. SyGuS instances are formulated in SyGuS-IF~\cite{RaghothamanU14}, a format built on top of SMT-LIB2~\cite{smtlib}.

We report here on the 4th SyGuS competition that took place in July 2017, in Heidelberg, Germany as a satellite event of CAV'17 (The 29th International Conference on Computer Aided Verification) and SYNT'17 (The Sixth Workshop on Synthesis). As in the previous competition there were four tracks: the general track, the conditional linear integer arithmetic track, the invariant synthesis track, and the programming by examples track. We assume most readers of this report are already familiar with the SyGuS problem and the tracks of SyGuS-Comp and thus refer the unfamiliar reader to the report on last year's competition~\cite{SyGuSComp15}. 

The report is organized as follows. Section~\ref{sec:benchs} describes the participating benchmarks. Section~\ref{sec:solvers}  lists the participating solvers, and briefly describes the main idea behind their strategy.  Section~\ref{sec:exp-set} provides details on the experimental setup. Section~\ref{sec:comp-results} gives an overview of the results per track. Section~\ref{sec:benchs-pres} provides details on the results, given from a single benchmark respective. Section~\ref{sec:discussion} concludes.


\section{Participating Benchmarks}
\label{sec:benchs}
In addition to last year's benchmarks, we received 4 new sets of benchmarks this year, which are shown in Table~\ref{tbl:new-benchmarks}.

\paragraph{Program Repair}
The 18 program repair benchmarks correspond to the task of generating small expression repairs that are consistent with a given set of input-output examples~\cite{repairbenchmarks}. These benchmarks were extracted from real-world Java bugs by manually analyzing the developer commits that involved changes to fewer than 5 lines of code. The key idea of the program repair approach is to the first localize the fault location in a buggy program and generate the corresponding input-output example behavior for the buggy expression from passing test cases. In the second phase, the task of repairing the buggy expression can be framed as a SyGuS problem, where the goal is to synthesize an expression that comes from a family of expressions defined using a context-free grammar of expressions and that satisfies the input-output example constraints.

\paragraph{Crypto Circuits}
The Crypto Circuits benchmarks comprise of tasks of synthesizing constant-time circuits that are cryptographically resilient to timing attacks~\cite{EldibWW16}. Consider a circuit $C$ with a set of \emph{private} inputs $I_0$ and a set of \emph{public} inputs $I_1$ such that if an attacker changes the values of
the public inputs and observes the corresponding output, she is unable to infer the values
of the private inputs (under standard assumptions about computational resources in cryptography). An attacker can gain information about private inputs by analyzing the time the circuit takes to compute the output values on public inputs, e.g. when a public input bit changes from 1 to 0, a specific output bit is guaranteed to
change from 1 to 0 independent of whether a particular private input bit is 0 or 1, but 
may change faster when this private input is 0, thus leaking information.
The timing attack can be prevented if the circuit satisfies the \emph{constant-time} property:
A constant-time circuit is the one in which the length of all input-to-output paths  measured in terms of number of gates
are the same.

The problem of synthesizing a new circuit $C'$ that is functionally equivalent to a given circuit $C$ such that $C'$ is a constant-time circuit can be formalized as a SyGuS problem. A context-free grammar can be used to define the set of all constant-time circuits with all input-to-output path lengths within a given bound, and the functional equivalence constraint can be expressed as a Boolean formula~\cite{EldibWW16}.

\paragraph{Instruction Selection}
The Instruction Selection benchmarks consist of tasks for synthesizing a ``Bit Test and Reset" instruction from the set of basic bitvector operations, in a way similar to the implementations supported by the x86 processors. These benchmarks comprise of 4 different addressing variants with increasing levels of complexity:
\begin{itemize}
	\item btr*: Read from register.
	\item btr-am-base*: Load from memory address base.
	\item btr-am-base-index*: Load from memory address base with indexing.
	\item btr-am-base-index-scale-disp*:  Load from memory address base with index shifted with scale.
\end{itemize}

\paragraph{Invariant Generation}
The invariant generation benchmarks comprise of the task of generating a loop invariant (as a conditional linear arithmetic expression) given the pre-condition, post-condition and the transition function corresponding to the loop body. The 7 new benchmarks~\cite{PadhiM17} correspond to loop invariant tasks adapted from several recent invariant inference papers including generating path invariants, abductive inference, and NECLA Static analysis benchmarks.

\begin{table}
	{\small{
			\begin{center}
				\scalebox{0.94}{
				\begin{tabular}{rcl}
					Benchmark Set &  \# of benchmarks &  Contributors \\ \hline \hline
					Invariant Generation & 7 & Saswat Padhi (UCLA)  \\
					Program Repair & 18 & 	Xuan Bach D Le (SMU), David Lo (SMU) and Claire Le Goues (CMU) \\
					Crypto Circuits & 214 & Chao Wang (USC) \\		
					Instruction Selection & 28 & Sebastian Buchwald (KIT) and Andreas Fried (KIT) \\
				\end{tabular}
			}
			\end{center}
			\caption{New Contributed Benchmarks}
			\label{tbl:new-benchmarks}
		}}
	\end{table}

\section{Participating Solvers}
\label{sec:solvers}
Six solvers were submitted to this year's competition. \eusolvernew, an improved version of \eusolver;  \cvcnew, an improved version of \cvc; \euphony, a solver built on top of \eusolver; \dryd, a solver specialized for conditional linear integer arithmetic;  \lig, a solver specialized for invariant generation problems; and \ethree, a solver specialized for the bitvector category of the PBE track, built on top of the enumerative solver.
Table~\ref{tbl:solvers-authors} lists the submitted solvers together with their authors, and Table~\ref{tbl:solvers-in-tracks} summarizes which solver participated in which track.

\begin{table}[b]
	{\small{
			\begin{center}
				\scalebox{0.94}{
				\begin{tabular}{r||l}
					Solver &  Authors \\ \hline \hline
					\eusolvernew  	& Arjun Radhakrishna (Microsoft) and
					Abhishek Udupa (Microsoft) \\
					\cvcnew 		& Andrew Reynolds (Univ. Of Iowa),
					Cesare Tinelli (Univ. of Iowa), and 
					Clark Barrett (Stanford)  \\
					\euphony        & Woosuk Lee (Penn), Arjun Radhakrishna (Microsft) and Abhishek Udupa (Microsoft) \\
					\dryd           & Kangjing Huang (Purdue Univ.), Xiaokang Qiu (Purdue Univ.), and Yanjun Wang (Purdue Univ.)\\
					\lig            & Saswat Padhi (UCLA) and Todd Millstein (UCLA)\\
					\ethree         & Ammar Ben Khadra (University of Kaiserslautern)			
					\\
				\end{tabular}}
			\end{center}
			\caption{Submitted Solvers }
			\label{tbl:solvers-authors}
		}}
\end{table}

\begin{table}[t]
	\begin{center}
		\begin{tabular}{r||rrrrrr}
			& \multicolumn{6}{c}{Solvers} \\
			Tracks & \rot{\eusolvernew} & \rot{\cvcnew} & \rot{\euphony} & \rot{\dryd} & \rot{\lig} & \rot{\ethree} \\ \hline \hline
			LIA         & 1 & 1 & 1 & 1 & 0 & 0 \\
			INV         & 1 & 1 & 1 & 1 & 1 & 0\\
			General     & 1 & 1 & 1 & 0 & 0 & 0\\ 
			PBE Strings & 1 & 1 & 1 & 0 & 0 & 0\\ 
			PBE BV      & 1 & 1 & 1 & 0 & 0 & 1
		\end{tabular}
	\end{center}
	\caption{Solvers participating in each track}
	\label{tbl:solvers-in-tracks}
\end{table}

 The \eusolvernew\ is based on the divide and conquer strategy~\cite{AlurCAV15}. The idea is to find different expressions that work correctly for different subsets of the input space, and unify them into a solution that works well for the entire space of inputs. The sub-expressions are typically found using enumeration techniques  and are then unified into the overall expression using machine learning methods for decision trees~\cite{AlurRU17}.

 The \cvcnew\ solver is based on an approach for program synthesis that is implemented inside an SMT solver~\cite{ReynoldsDKTB15}. This approach extracts solution functions from unsatisfiability proofs of the negated form of synthesis conjectures, and uses  counterexample-guided techniques for quantifier instantiation (CEGQI) that make finding such proofs practically feasible. \cvcnew\ also combines enumerative techniques, and symmetry breaking techniques~\cite{ReynoldsT17}. 
 
 The \euphony\ solver leverages statistical program models to accelerate the \eusolver. The underlying statistical model is called probabilistic higher-order grammar (PHOG), a generalization of probabilistic context-free grammars (PCFGs). The idea is to use existing benchmarks and the synthesized results to learn a weighted grammar, and give priority to candidates which are more likely according to the learned weighted grammar.

 The \dryd\ solver combines enumerative and symbolic techniques. It considers benchmarks in conditional linear integer arithmetic theory (LIA), and can therefore assume all have a solution in some pre-defined decision tree normal form. It then tries to first get the correct height of a normal form decision tree, and then tries to synthesize a solution of that height. It makes use of parallelization, using as many cores as are available, and of optimizations based on solutions of typical LIA SyGuS problems.  
 
 The \lig\ solver~\cite{PadhiM17} for invariant synthesis extends the data-driven approach to inferring sufficient loop invariants from a collection of program states~\cite{PadhiSM16}. Previous approaches to invariant synthesis were restricted to using a fixed set, or a fixed template for features, e.g., ICE-DT~\cite{ICEDT,GNMR16} requires the shape of constraints (such as octagonal) to be fixed apriori. Instead \lig\, starts with no initial features, and automatically grows the feature set as necessary using program synthesis techniques. It reduces the problem of loop invariant inference to a series of precondition inference problems and uses a Counterexample-Guided Inductive Synthesis (CEGIS) loop to revise the current candidate.
 
 The \ethree\ solver for PBE bitvector programs, is built on top of the enumerative solver~\cite{AlurBJMRSSSTU13,UdupaRDMMA13}. It improves on the original \enum\ solver by applying unification techniques~\cite{AlurCAV15} and avoiding calling an SMT solver, since on PBE tracks there are no semantic constraints other than the input-to-output examples which can be checked without invoking an SMT solver.

\label{sec:results}

\section{Experimental Setup} 
\label{sec:exp-set}
The solvers were run on the StarExec platform~\cite{starexec} with a dedicated cluster of 12 nodes, where each node
consisted of two 4-core 2.4GHz Intel processors with 256GB RAM and a 1TB hard drive. The memory
usage limit of each solver run was set to 128GB. The wallclock time limit was set to 3600 seconds (thus,
a solver that used all cores could consume at most 14400 seconds cpu time).
The solutions that the solvers produce are being checked for both syntactic and semantic correctness.
That is, a first post-processor checks that the produced expression adheres to the grammar specified in
the given benchmark, and if this check passes, a second post-processor checks that the solution adheres
to semantic constraints given in the benchmark (by invoking an SMT solver).

\section{Results Overview}
\label{sec:comp-results}
\begin{figure}
	\centering
	\includegraphics[scale=0.5]{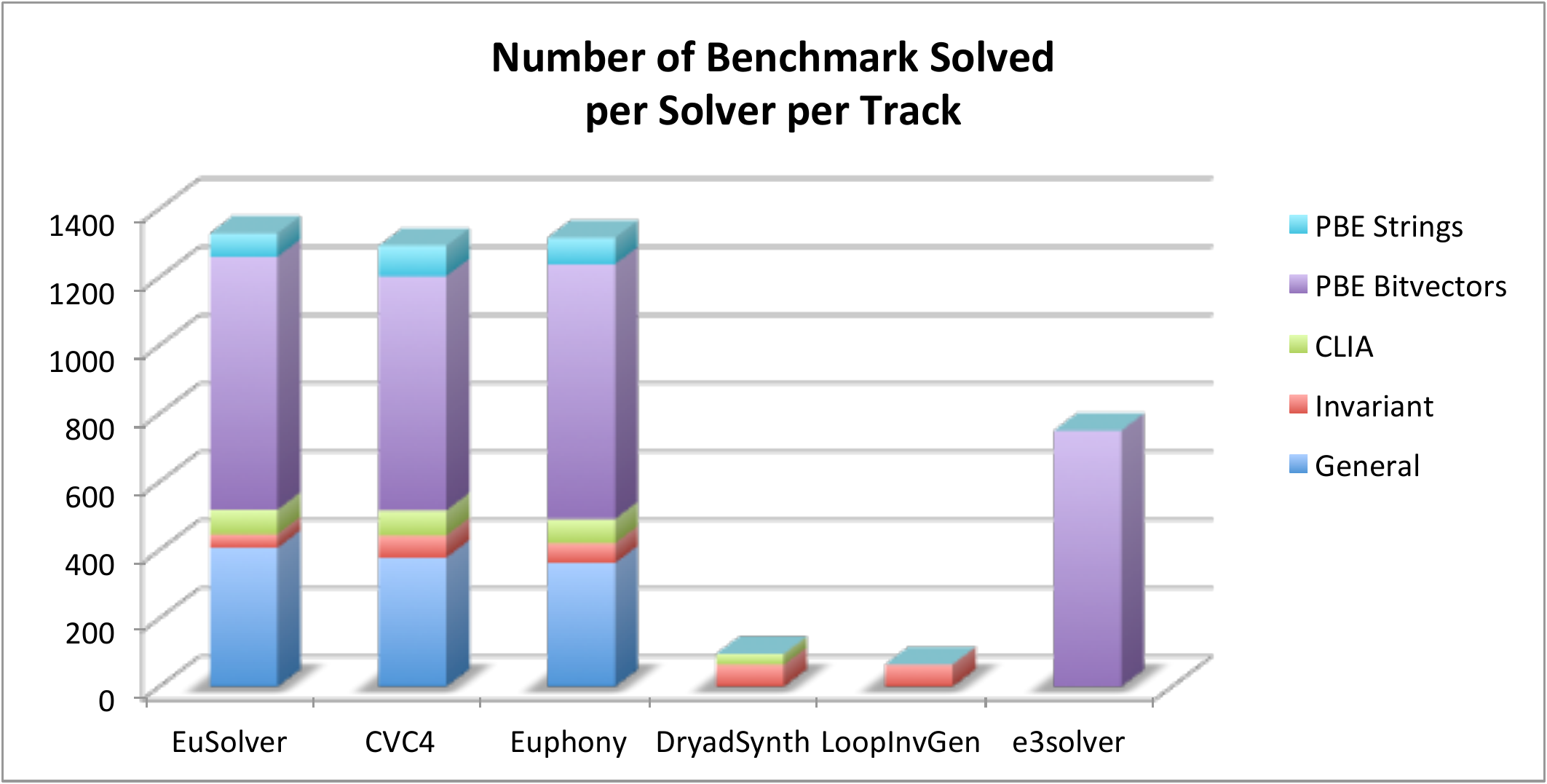}
	\caption{The overall combined results for each solver on benchmarks from all five tracks.}
	\label{fig:combinedresults}
\end{figure}

The combined results for all tracks are given in Figure~\ref{fig:combinedresults}. The figure shows the sum of benchmarks solved by the solvers for each track. We can observe that the $\eusolvernew$ solved the highest number of benchmarks in the combined tracks, and   the \euphony\ solver and the $\cvcnew$ solver  solved almost as many.

The primary criterion for winning a track was the number of benchmarks solved, but we also analyzed the time to solve and the the size of the generated expressions. Both where classified using a pseudo-logarithmic scale as follows.
For time to solve the scale is [0,1), [1,3), [3,10), [10,30),[30, 100), [100,300), [300, 1000), [1000,3600), $>$3600. That is the first ``bucket'' refers to termination in less than one second, the second to termination in one to three second and so on. We say that a solver solved a certain benchmark \emph{among the fastest} if the time it took to solve that benchmark was on the same bucket as that of the solver who solved that benchmark the fastest. 
For the expression sizes the pseudo-logarithmic scale we use is [1,10), [10,30), [30,100), [100,300), [300,1000), $>$1000 where expression size is the number of nodes in the SyGuS parse-tree.
In some tracks there was a tie or almost a tie in terms of the number of solved benchmarks, but the differences in the time to solve where significant.
We also report on the number of benchmarks \emph{solved uniquely} by a solver (meaning the number of benchmark that solver was the single solver that managed to solve them).

\begin{figure}
	\begin{center}
		\begin{minipage}{1\textwidth}
			\centering
			\includegraphics[scale=0.9,width=1\textwidth]{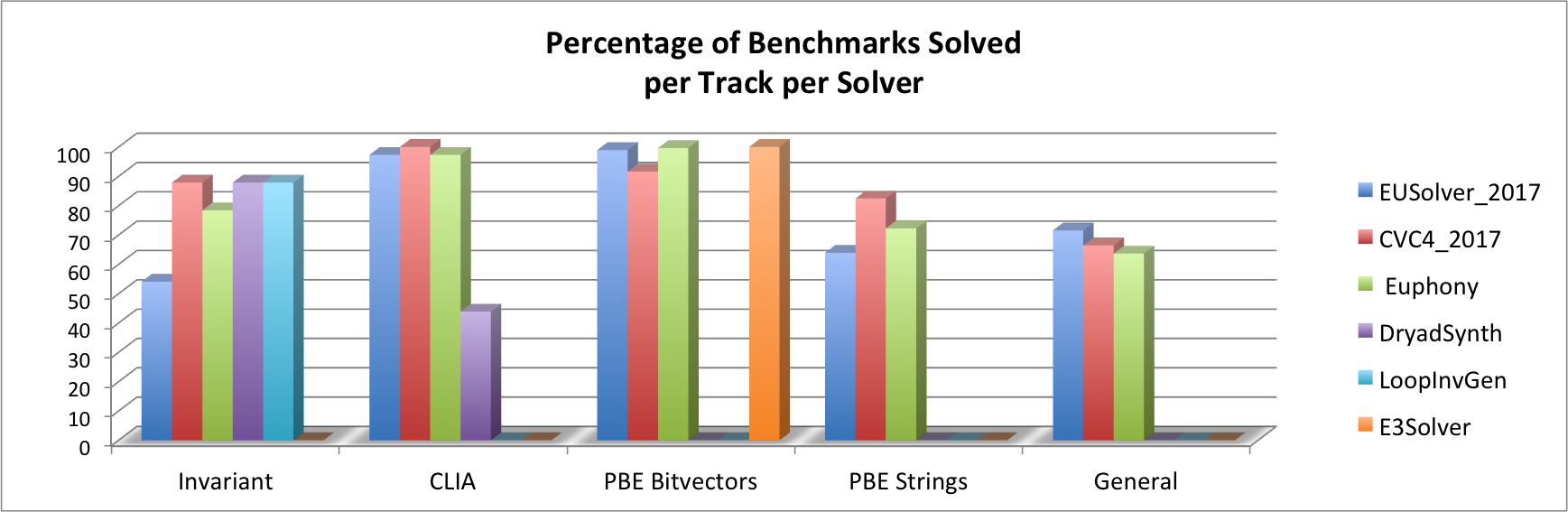}
		\end{minipage}
		\\
		\vspace{2mm}
		\begin{minipage}{1\textwidth}
			\centering
			\includegraphics[scale=0.9,width=1\textwidth]{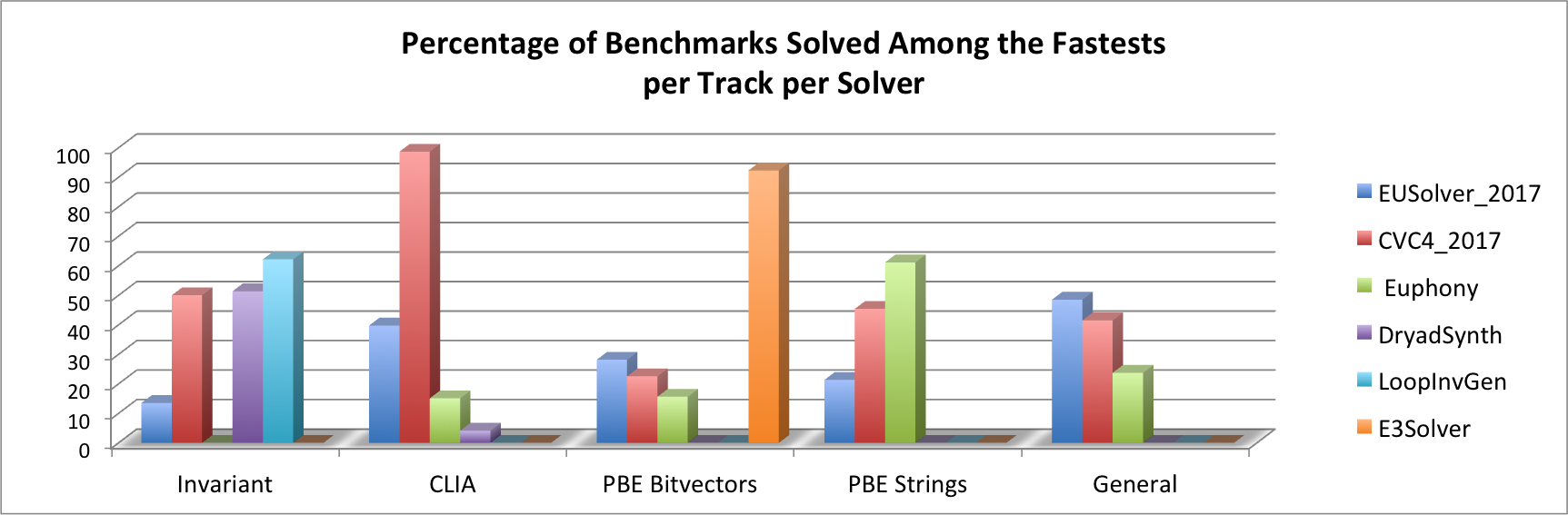}
		\end{minipage}
	\end{center}
	\caption{Percentage of benchmarks solved by the different solvers across all tracks, and the percentage of benchmarks a solver solved among the fastest for that benchmarks (according to the logarithmic scale)}\label{fig:resultsPerTrack}	
\end{figure}

Figure~\ref{fig:resultsPerTrack} shows the percentage of benchmarks solved by each of the solvers in each of the tracks (in the upper part) and the number of benchmarks solved among the fastest by each of the solvers in each of the tracks (in the lower part) and the number of benchmarks solved among the fastest. 

\paragraph{General Track}
In the general track the \eusolvernew\ solved more benchmarks than all others (407), the \cvcnew\ came second, solving 378 benchmarks, and \euphony\ came third, solving 362 benchmarks. The same order appears in the number of benchmarks solved among the fastest: \eusolvernew\ with 276, \cvcnew\ with 236, and \euphony\ with 135. In terms of benchmarks solved uniquely by a solver, we have that \eusolvernew\ solved 34 uniquely, \cvcnew\ solved 9 uniquely, and \euphony\ solved 2 uniquely.

\begin{table}[t]
	\begin{center}
		\scalebox{0.9}{
		\begin{tabular}{lr||rrrrrrrrrrrr|r}
			 &	& \rot{Compiler Optimizations and Bit Vectors}	& \rot{Let and Motion Planning} &	\rot{Invariant Generation with Bounded Ints} &	\rot{Invariant Generation with Unbounded Ints} &	\rot{Multiple Functions}	& \rot{Arrays} &	\rot{Hackers Delight} &	\rot{Integers} &	\rot{Program Repair} &	\rot{ICFP} &	\rot{Cryptographic Circuits} &	\rot{Instruction Selection} & {Total}\\\hline \hline
\multicolumn{2}{l||}{Number of benchmarks}  & 32 & 30 & 28 & 28 & 32 & 31 & 44 & 34 & 18 & 50 & 214 & 28 & 569 \\ \hline			 
\multirow{3}{*}{Solved} & \eusolvernew\ &	16	& 10	& 24	&24 &	18	& 31	& 35 &	33 &	14 &	50& 	152& 	0 & 407 \\
& \cvcnew\ &	15	& 15	& 24	& 24	& 12	& 31 &	44& 	34&	14&	48&	117&	0 & 378 \\
& \euphony\	& 19	&10 &	24 &	24 &	18	& 31	& 44	& 33	& 14	& 50	& 95 &	0 & 362 \\ \hline
\multirow{3}{*}{Fastest} & \eusolvernew\ &	7	&2 	& 12	& 14	& 6	& 5	& 20	& 14 &	13 &	40	& 143 &	0 & 276\\
 & \cvcnew &	11 &	15 &	18	& 19	& 9	& 31 &	44	& 33	& 7& 	19	& 30	& 0 & 236 \\
& \euphony	& 16 &	2	& 8	& 13 &	13 &	4	& 27 &	14 &	9	& 29	& 0& 	0 & 135 \\ \hline
\multirow{3}{*}{Uniquely} & \eusolvernew\ &	0	& 0& 	0	& 0& 	0	& 0& 	0	& 0 &	0	& 0	& 34 &	0 & 34 \\
& \cvcnew &	1	& 5&	0&	0&	1&	0	&0&	1&	1&	0	&0&	0 & 9\\
& \euphony	& 2	& 0	& 0& 	0	& 0& 	0	& 0& 	0	& 0& 	0 &	0	& 0 & 2\\ \hline			
		\end{tabular}}
	\end{center}
	\caption{Solvers performance across all categories of the general track}
	\label{tbl:general-categories}
\end{table}

We partition the benchmarks of the general track according to categories where different categories consists of related benchmarks. The results per category are given in the Table~\ref{tbl:general-categories}. We can see that \eusolvernew\ preformed better than others in the categories of program repair, icfp and cryptographic circuits. The \cvcnew\ solver preformed better than others in the categories of let and motion planning, invariant generation with bounded and unbounded integers, arrays, integers and hacker's delight. The \euphony\ solver preformed better than others in the categories of  multiple functions, compiler optimizations and bitvectors. We can also observe that none of the solvers could solve any of the instruction selection benchmarks.

\begin{figure}
	\begin{center}
		\begin{minipage}{1\textwidth}
			\centering
			\includegraphics[scale=0.9,width=1\textwidth]{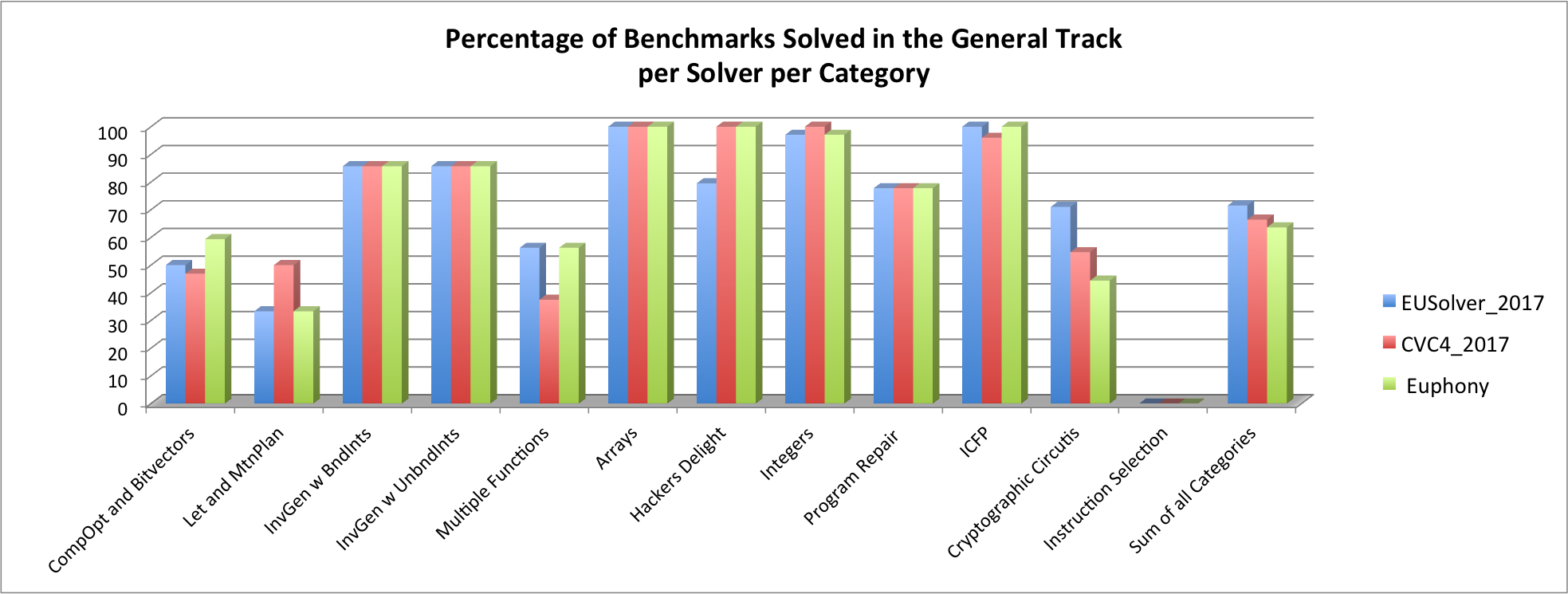}
		\end{minipage}
		\\
		\vspace{2mm}
		\begin{minipage}{1\textwidth}
			\centering
			\includegraphics[scale=0.9,width=1\textwidth]{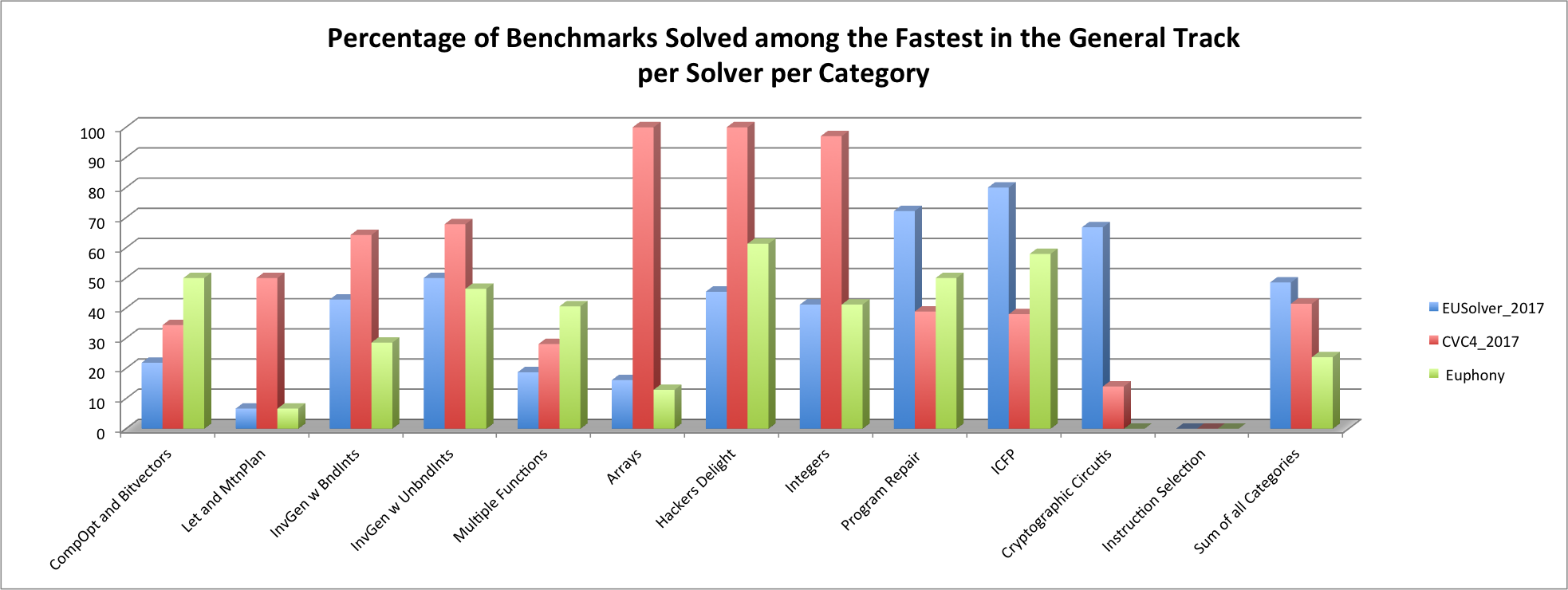}
		\end{minipage}
	\end{center}
	\caption{Percentage of benchmarks solved by the different solvers across all categories of the general track, and the percentage of benchmarks a solver solved among the fastest for that benchmark (according to the logarithmic scale).  }	
\end{figure}

\paragraph{Conditional Linear Arithmetic Track}
In the CLIA track the \cvcnew\ solved all 73 benchmarks, \euphony\ and \eusolvernew\ solved 71 benchmarks, and \dryd\ solved 32 benchmarks. The \cvcnew\ solver solved 72 benchmarks among the fastest, followed by \eusolvernew\ which solved 29 among the fastest, \euphony\ which solved 11 among the fastetst, and \dryd\ which solved 3 among the fastest. None of the benchmarks where solved uniquely.

\paragraph{Invariant Generation Track}
In the invariant generation  track, both the \lig\ solver and the \cvcnew\ solver solved 65 out of 74 benchmarks, the \dryd\ solver solved 64 benchmarks, the Euphony solver solved 48 benchmarks and EUSolver solved 40 benchmarks. In terms of the time to solve the differences where more significant. The \lig\ solver solved 54 benchmarks among the fastest, followed by \cvcnew\ which solved 38 among the fastest, and \dryd\
which solved 36 among the fastest. There was one benchmark that only one solver solved, this is the \texttt{hola07.sl} benchmark, and the solver is \lig.

\paragraph{Programming By Example BV Track}
In the PBE track using BV theory, the \ethree\ solver solved all 750 benchmarks, \euphony\ solved 747 benchmarks, \eusolvernew\ solved 242, benchmarks, and \cvcnew\ solved 687 benchmarks. The \ethree\ solver solved 692 among the fastest, \eusolvernew\ solved 211 among the fastest, \cvcnew\ solved 169 among the fastest, and \euphony\ solved 117 among the fastest. Three benchmarks where solved uniquely by \ethree, these are: \texttt{13_1000.sl}, \texttt{40_1000.sl} and \texttt{89_1000.sl}.

\paragraph{Programming By Example Strings Track}
In the PBE track using SLIA theory, the \cvcnew\ solved 89 out of 108 benchmarks, \euphony\ solved 78, and \eusolvernew\ solved 69. The \euphony\ solver solved 66 benchmarks among the fastest, \cvcnew\ solved 49 among the fastest and \eusolvernew\ solved 23 among the fastest. Nine benchmarks where solved by only one solver, which is \cvcnew.

\section{Detailed Results}
\label{sec:benchs-pres}
In the following section we show the results of the competition from the benchmark's perspective. 
For a given benchmark we would like to know: how many solvers solved it, what is the min and max time to solve,  what are the min and max size of the expressions produced, which solver solved the benchmark the fastest, and which solver produced the smallest expression.

\begin{figure*}
\noindent\makebox[\textwidth]{
\scalebox{0.6}{
	\begin{tabular}{c}
		\includegraphics[width=9.0in,bb=7 9 923 476]{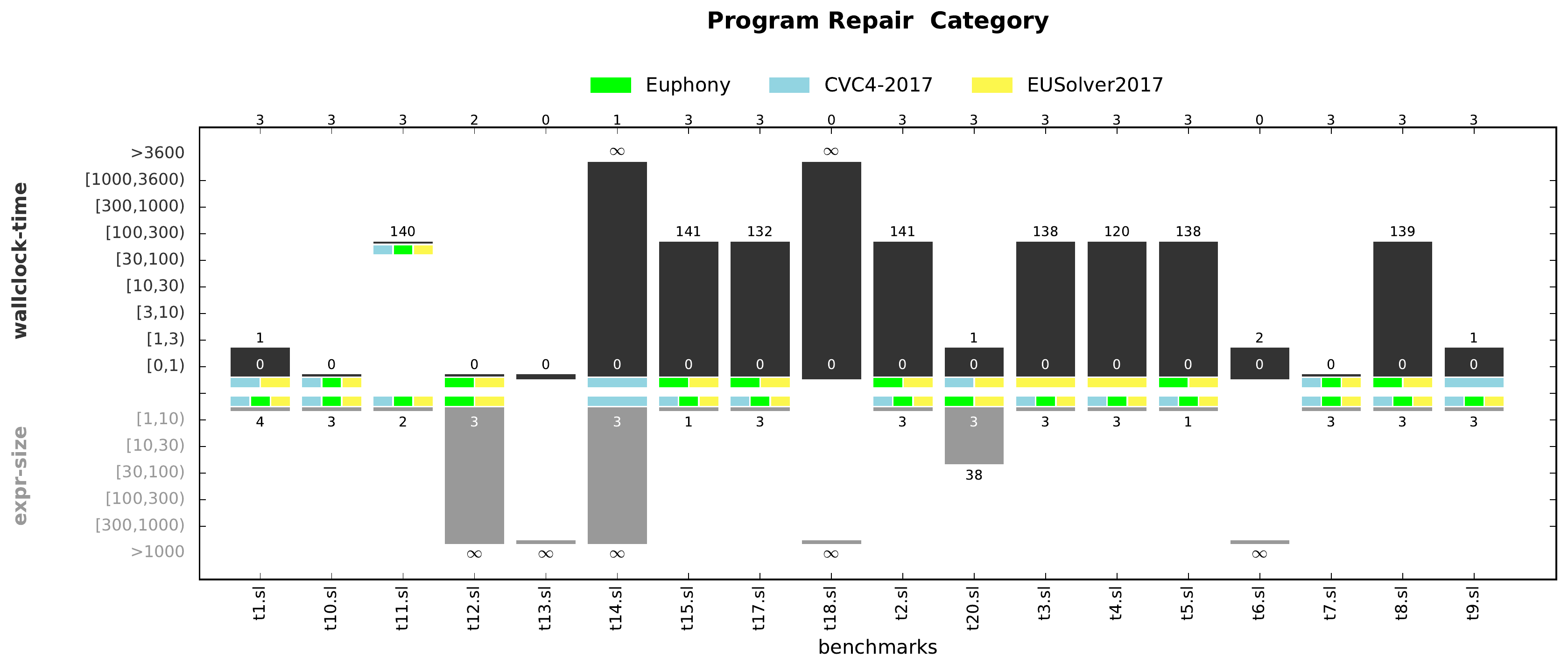} \\[3cm]
		\includegraphics[width=9.7in,bb=7 9 925 460]{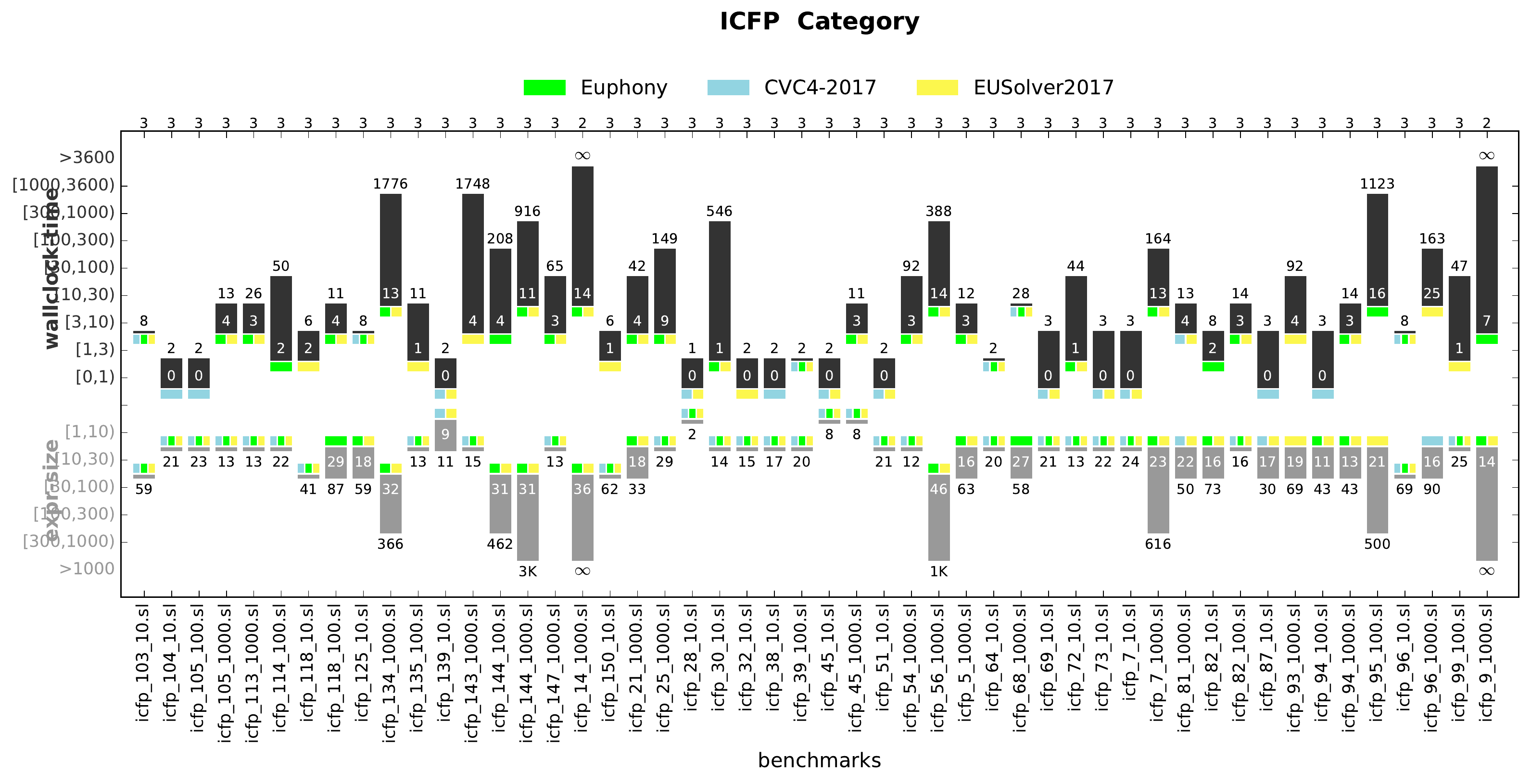} 
	\end{tabular}
}}
\caption{Evaluation of Program Repair and ICFP Categories of the General Track.}\label{fig:prog-rep-icfp}
\end{figure*}

We represents the results per benchmark in groups organized per tracks and categories. For instance, Fig.~\ref{fig:prog-rep-icfp} at the top presents details of the program repair benchmarks. The black bars show the range of the time to solve among the different solvers in pseudo logarithmic scale (as indicated on the upper part of the y-axis). Inspect for instance benchmark \texttt{t\_2.sl}. The black bar indicates that the fastest solver to solve it used less than 1 second, and the slowest used between 100 to 300 seconds. 
The black number above the black bar indicates the exact number of seconds (floor-rounded to the nearest second) it took the slowest solver to solve a benchmark (and $\infty$ if at least one solver exceeded the time bound). Thus, we can see that the slowest solver to solve \texttt{t\_2.sl} took 141 seconds to solve it. The white number at the lower part of the bar indicates the time of the fastest solver to solve that benchmark. Thus, we can see that the fastest solver to solve \texttt{t\_2.sl} required less than 1 second to do so. The colored squares/rectangles next to the lower part of the black bar, indicate which solvers were the fastest to solve that benchmark (according to the solvers' legend at the top). Here, \emph{fastest} means in the same logarithmic scale as the absolute fastest solver. For instance, we can see that \euphony\ and \eusolvernew\ were the fastest to solve \texttt{t\_2.sl}, solving it in less than a second
and that among the 2 solvers that solved \texttt{t\_3.sl} only \eusolvernew\ solved it in less than 1 seconds. 

Similarly, the gray bars indicate the range of expression sizes in pseudo logarithmic scales (as indicated on the lower part of the y-axis), where the size of an expression is determined by the number of nodes in its parse tree.
The black number at the bottom of the gray bar indicates the exact size expression of the largest solution (or $\infty$ if it exceeded 1000), and the white number at the top of the gray bar indicates the exact size expression of the smallest solution (when the smallest and largest size of expressions are in the same logarithmic bucket (as is the case in \texttt{t\_2.sl}), we provide only the largest expression size, thus there is no white number on the gray bar). The colored squares/rectangles next to the upper part of the gray bar indicates which solvers (according to the legend) produced the smallest expression (where \emph{smallest} means in the same logarithmic scale as the absolute smallest expression). For instance, for \texttt{t\_20.sl} the smallest expression produced had size 3, and 2 solvers out of the 3 who solved it managed to produce an expression of size less than 10.  

Finally, at the top of the figure above each benchmark there is a number indicating the number of solvers that solved that benchmark. For instance, one solver solved \texttt{t\_14.sl}, two solvers solved \texttt{t\_12.sl}, three solvers solved \texttt{t\_2.sl}, and no solver solved \texttt{t\_6.sl}. Note that the reason \texttt{t\_6.sl} has 2 as the upper time bound, is that that is the time to terminate rather than the time to solve. Thus, all solvers aborted within less than 2 seconds, but either they did not produce a result, or they produced an incorrect result. When no solver produced a correct result, there are no colored squares/rectangles next to the lower parts of the bars, as is the case for \texttt{t\_6.sl}.

\begin{figure*}
\noindent\makebox[\textwidth]{
	\scalebox{0.6}{
		\begin{tabular}{c}
			\includegraphics[width=9.5in,bb=7 9 923 476]{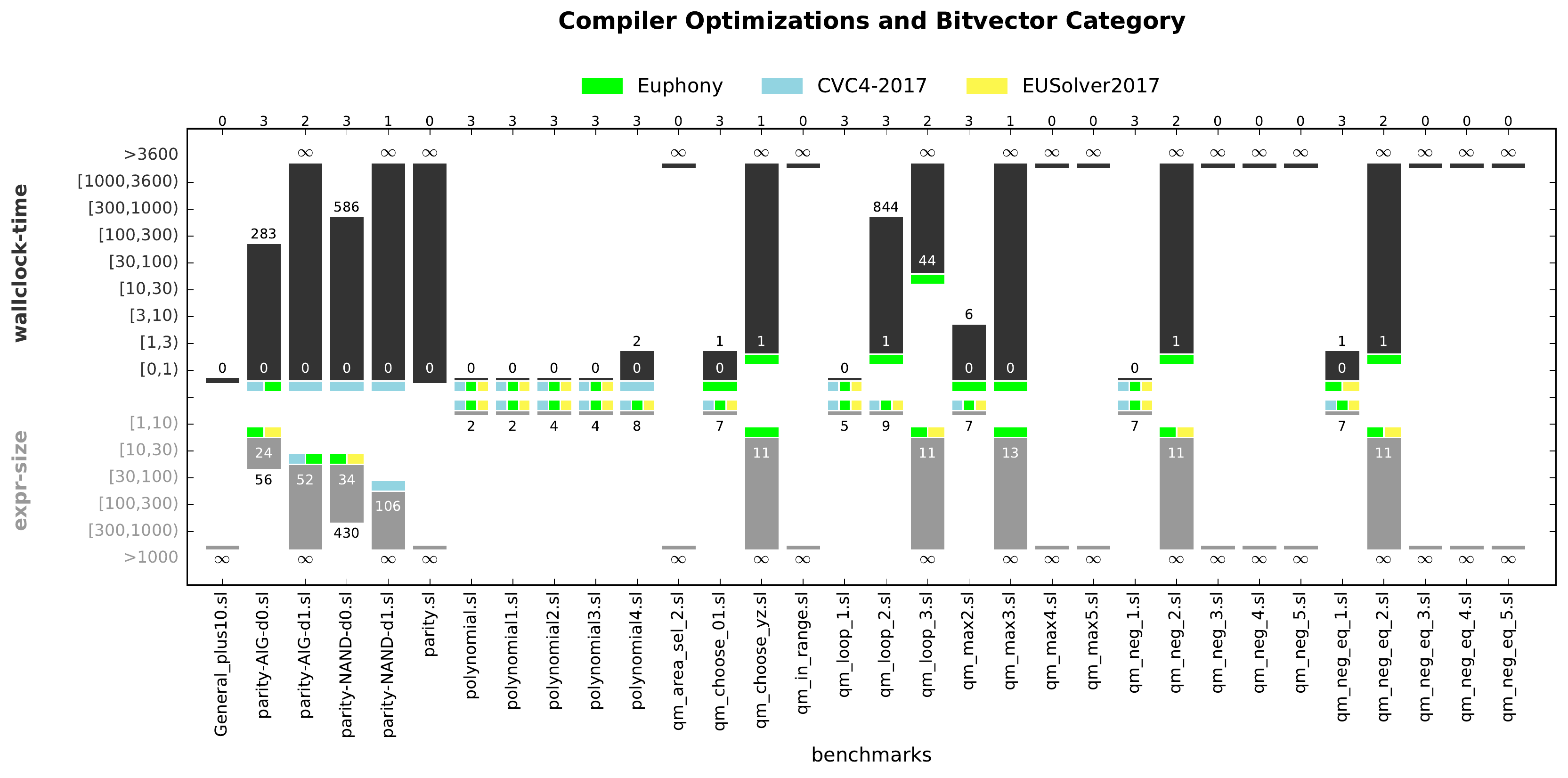} \\[3cm]
			\includegraphics[width=9.5in,bb=7 9 925 460]{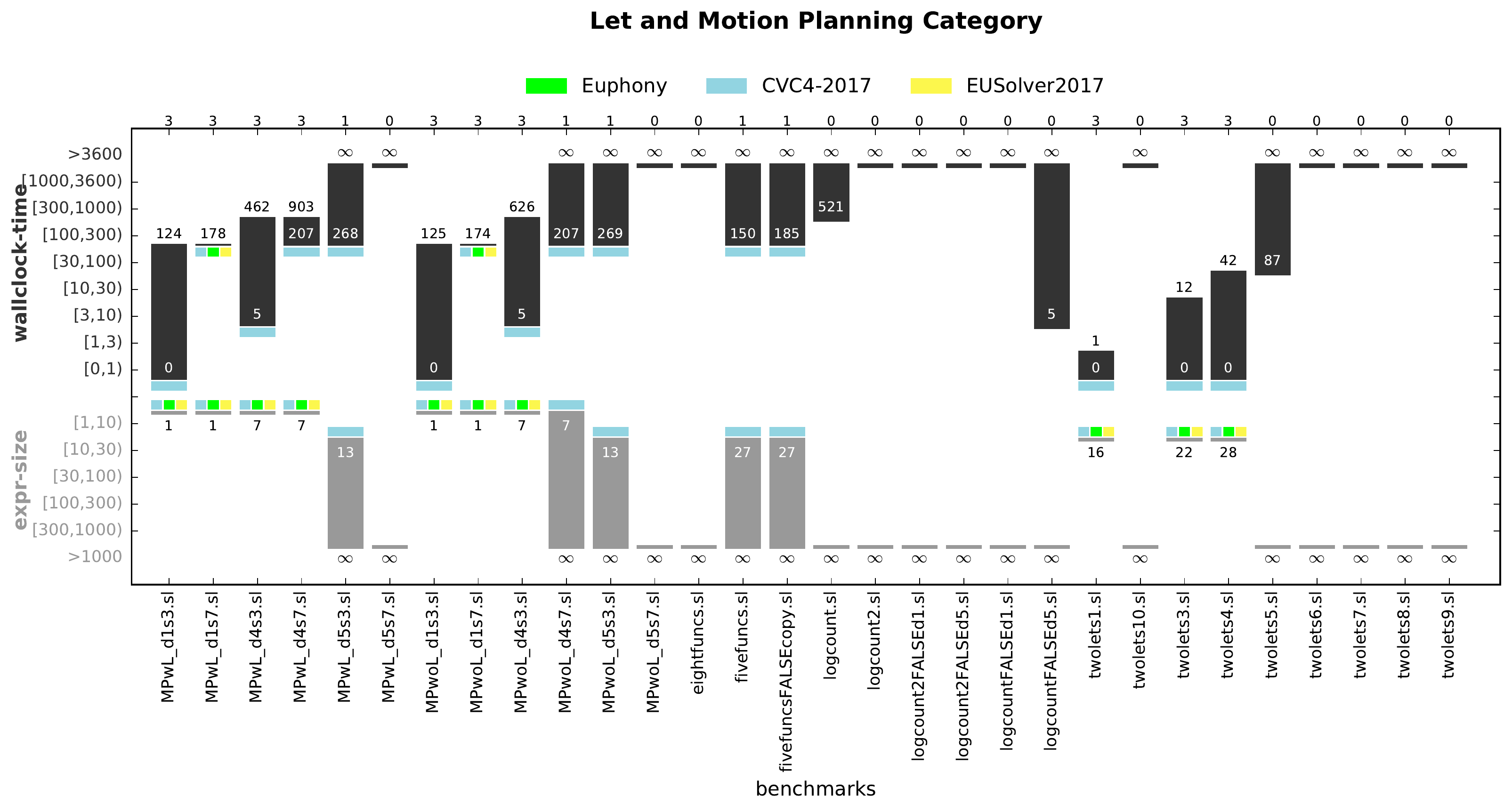} 
		\end{tabular}
	}}
	\caption{Evaluation of Compiler Optimizations, Bitvectors, Let and Motion Planning Categories of the General Track.}\label{fig:let-mot-plan}
\end{figure*}

\begin{figure*}
\noindent\makebox[\textwidth]{
	\scalebox{0.6}{
		\begin{tabular}{c}
			\includegraphics[width=9.5in,bb=7 9 923 476]{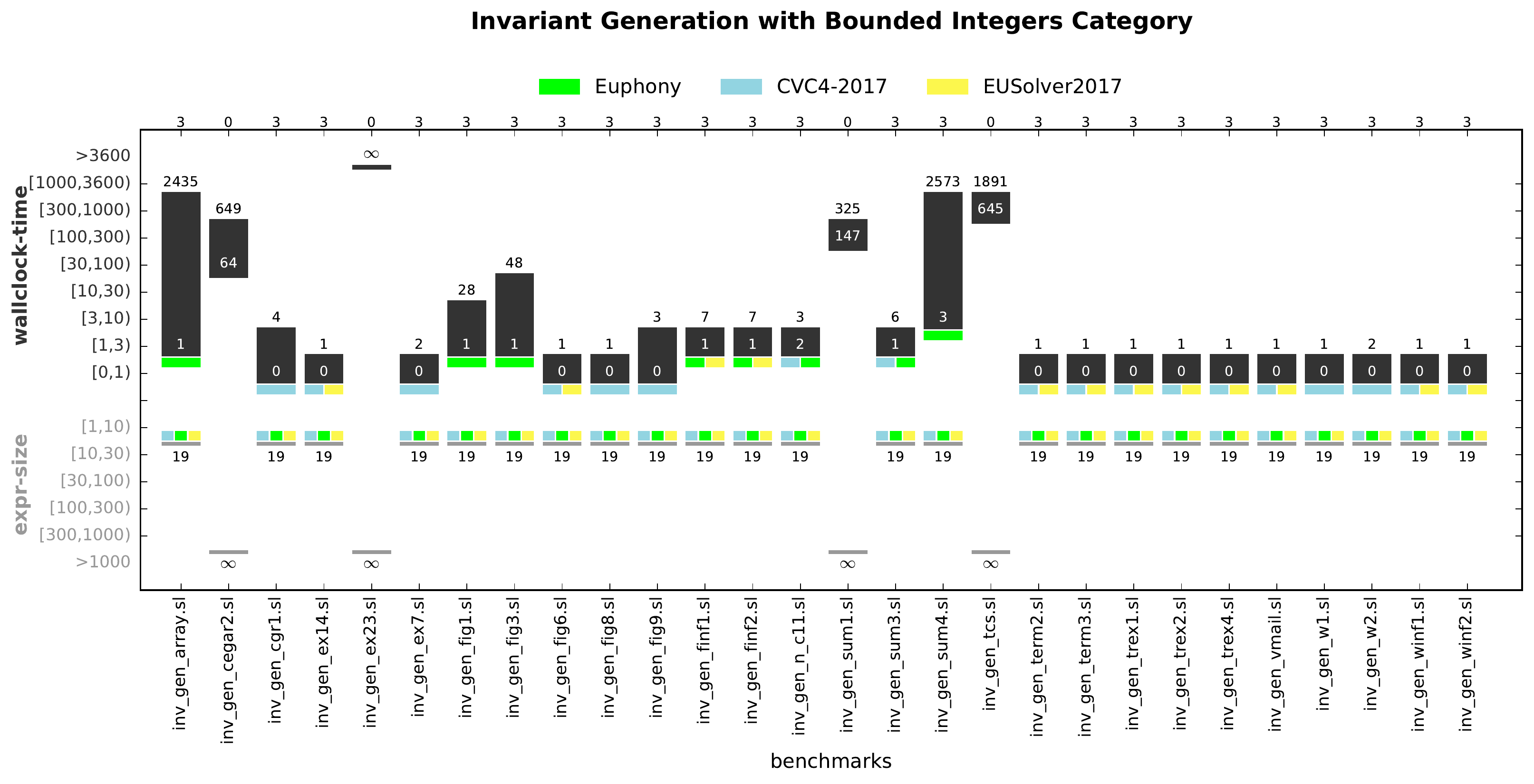} \\[3cm]
			\includegraphics[width=9.5in,bb=7 9 925 460]{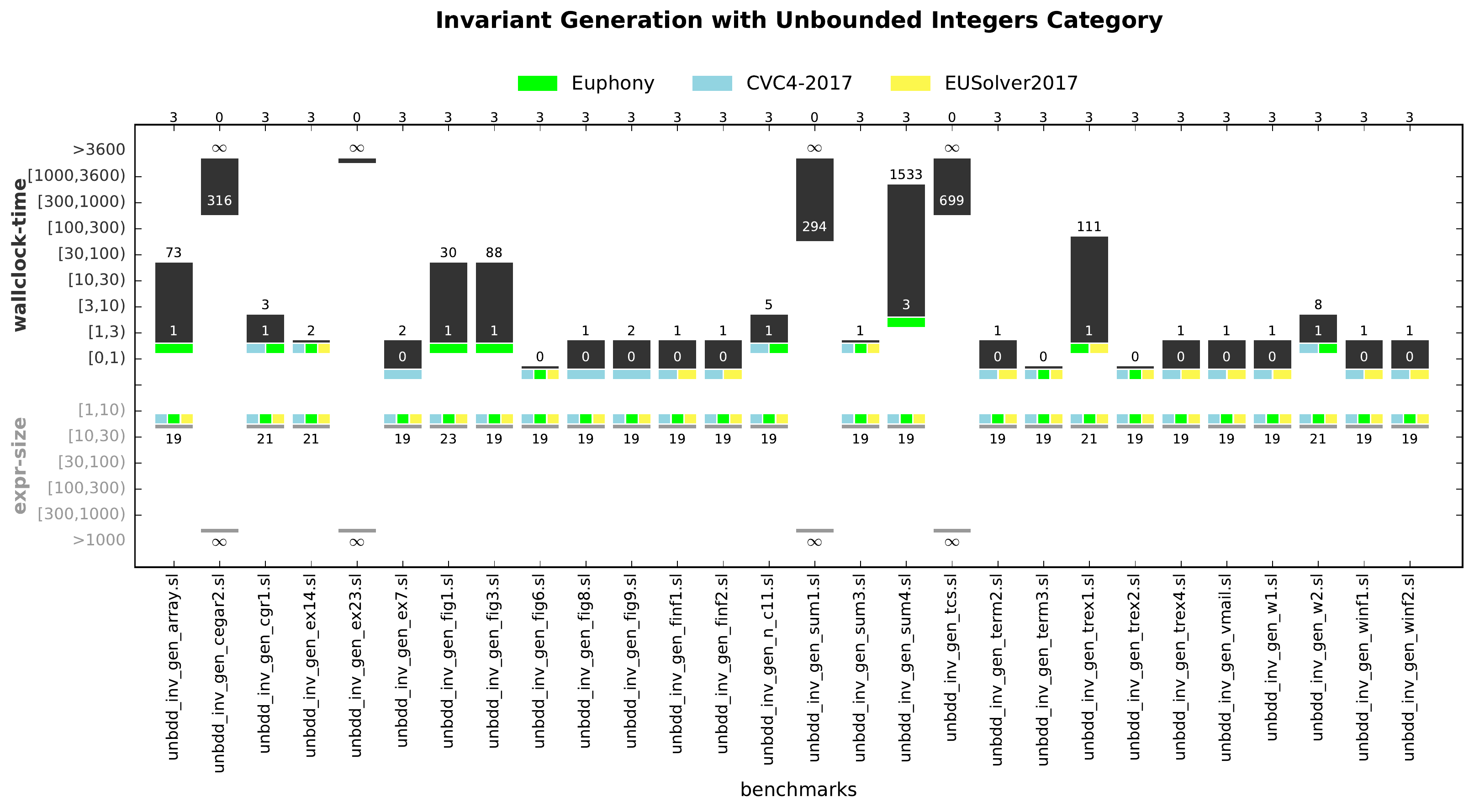} 
		\end{tabular}
	}}
	\caption{Evaluation of Invariant Category of the General Track.}\label{fig:inv-results}
\end{figure*}

\begin{figure*}
\noindent\makebox[\textwidth]{
\scalebox{0.6}{
	\begin{tabular}{c}
		\includegraphics[width=9.5in,bb=7 9 923 476]{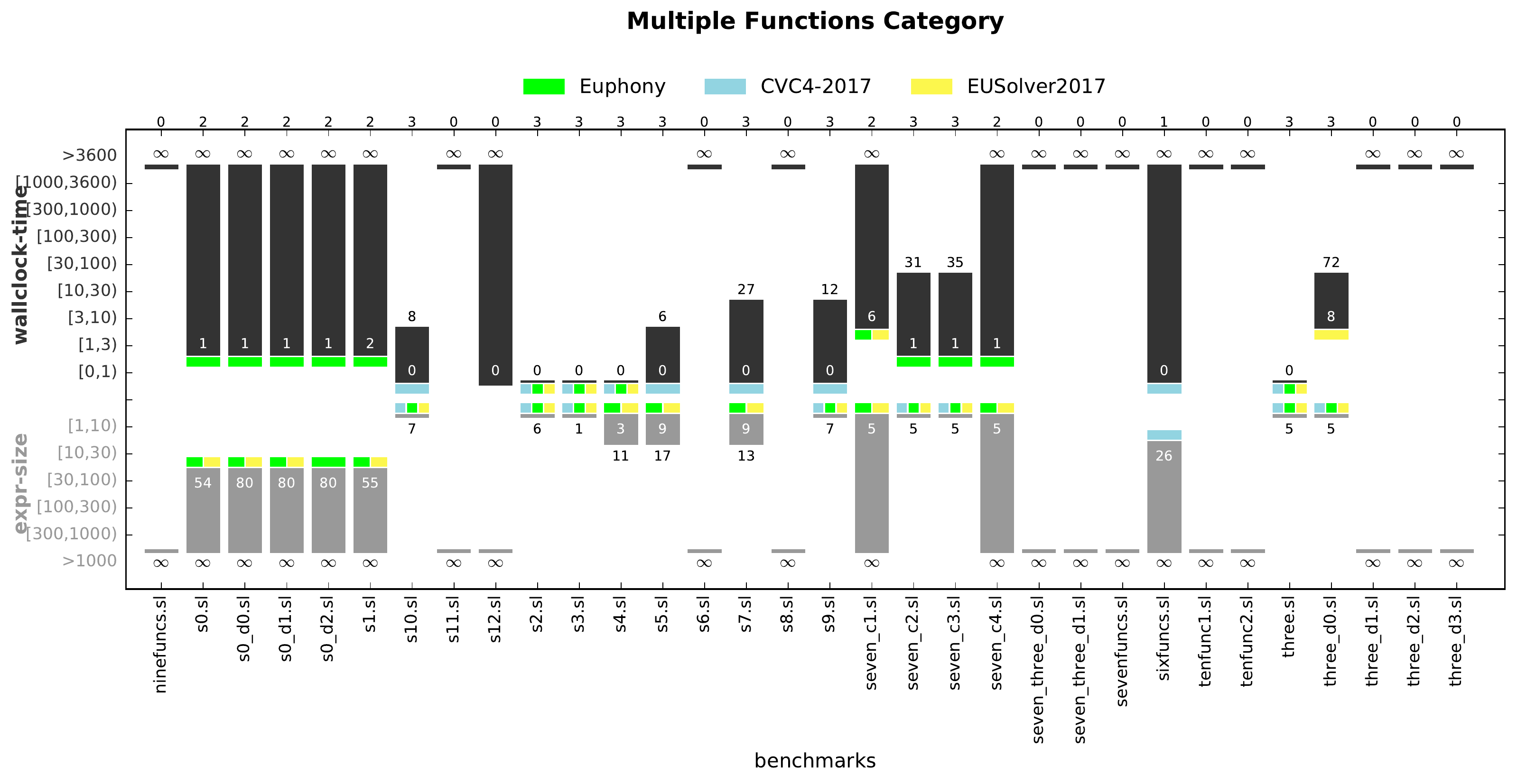} \\[3cm]
		\includegraphics[width=9.5in,bb=7 9 925 460]{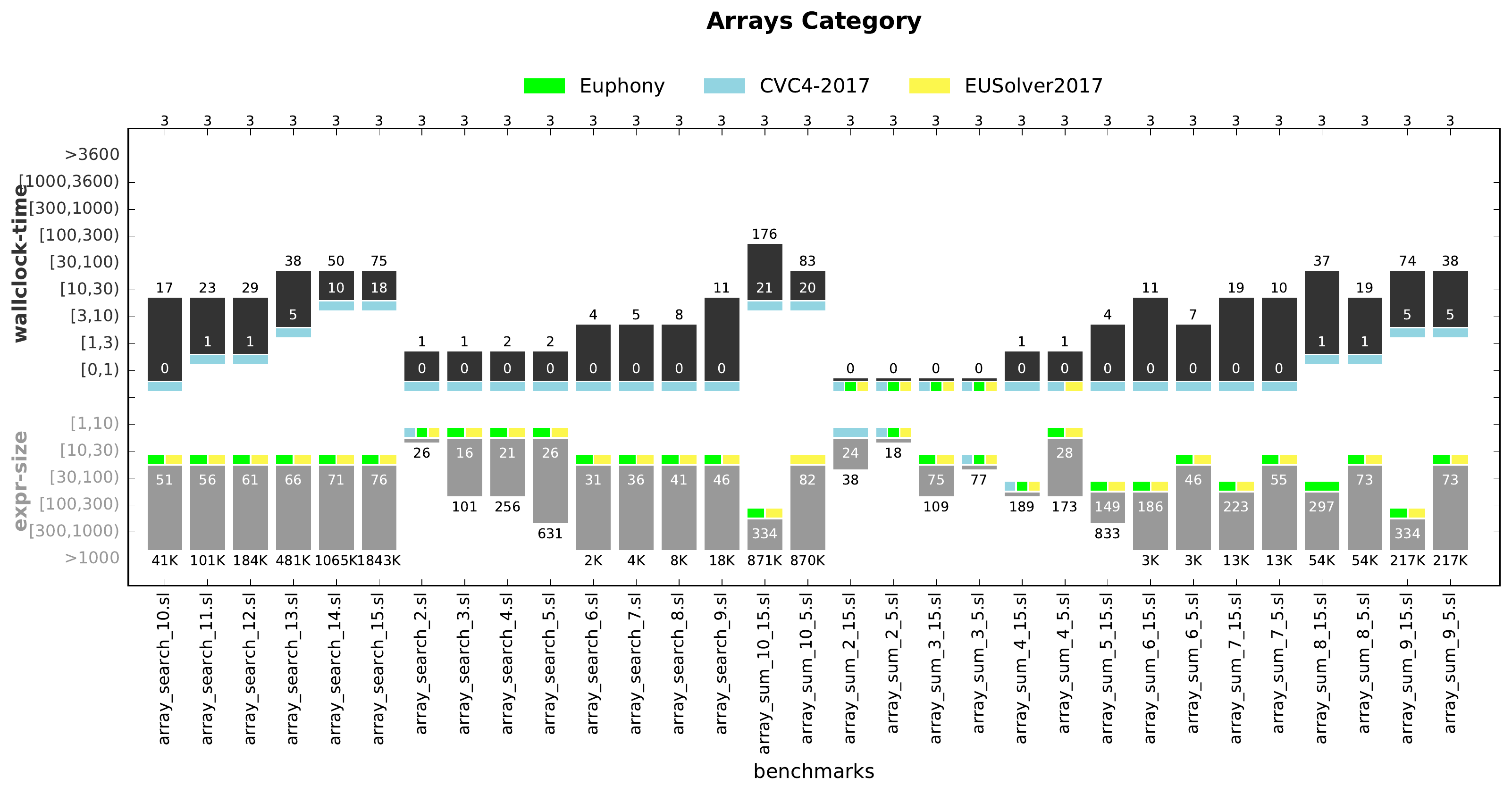} 
	\end{tabular}
}}
\caption{Evaluation of Multiple Functions and Arrays Categories of the General Track.}\label{fig:mult-func-arr}
\end{figure*}

\begin{figure*}
\noindent\makebox[\textwidth]{
	\scalebox{0.6}{
		\begin{tabular}{c}
			\includegraphics[width=9.5in,bb=7 9 923 476]{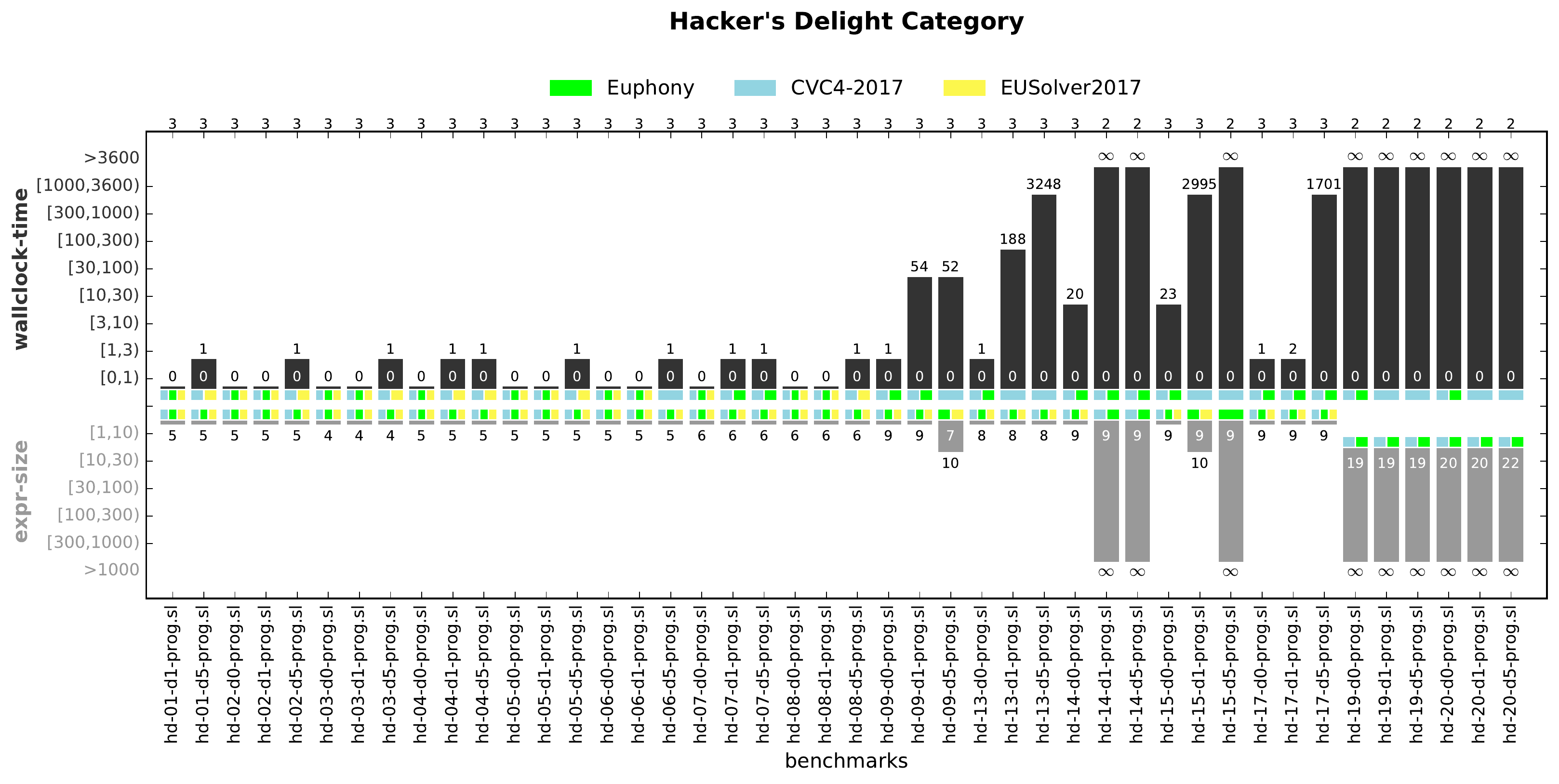} \\[3cm]
			\includegraphics[width=9.5in,bb=7 9 925 460]{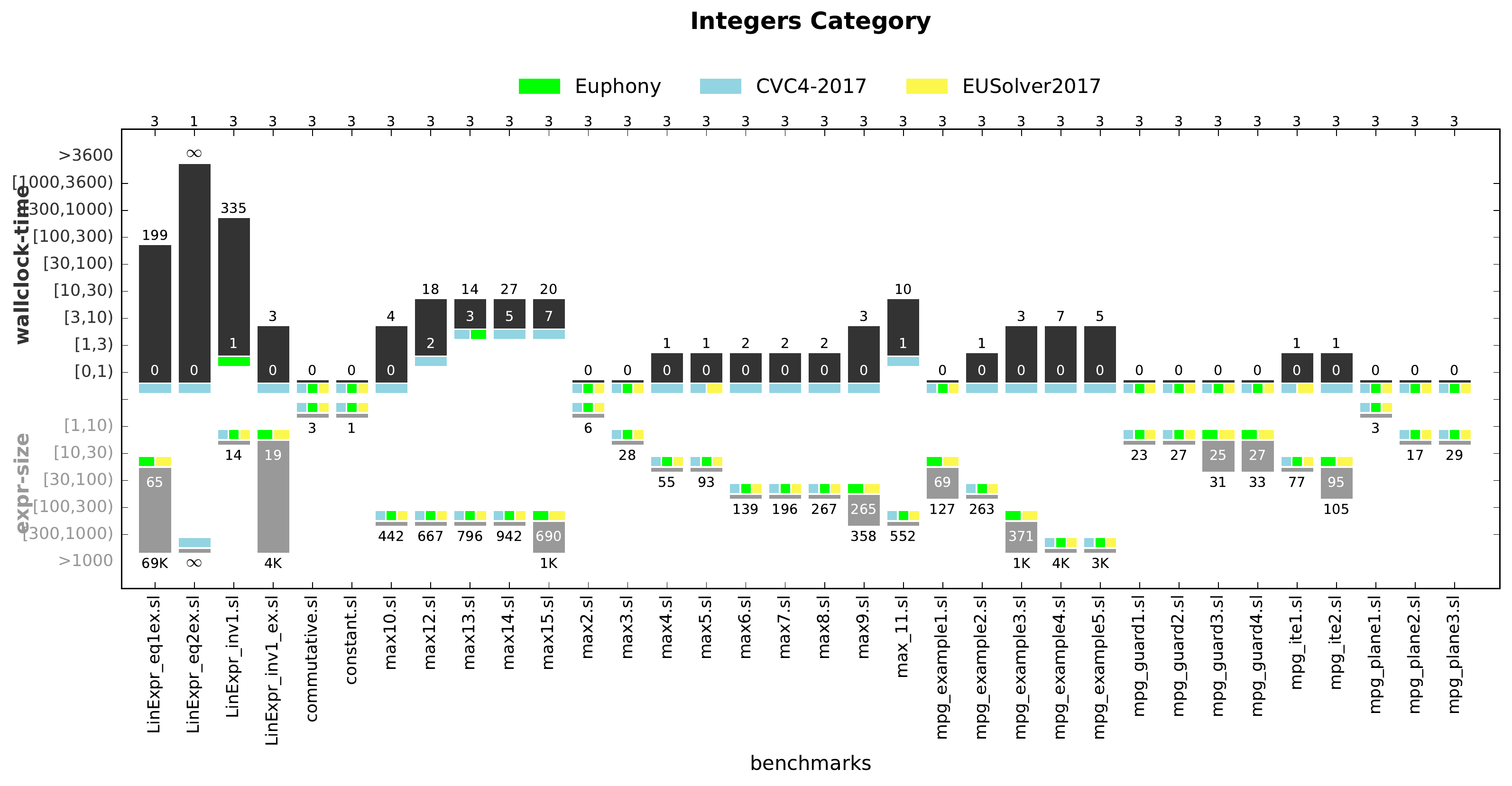} 
		\end{tabular}
	}}
	\caption{Evaluation of Hacker's Delight and Integers Categories of the General Track.}\label{fig:hd-int}
\end{figure*}

\begin{figure*}
\noindent\makebox[\textwidth]{
	\scalebox{0.6}{
		\begin{tabular}{c}
			\includegraphics[width=9.5in,bb=7 9 923 476]{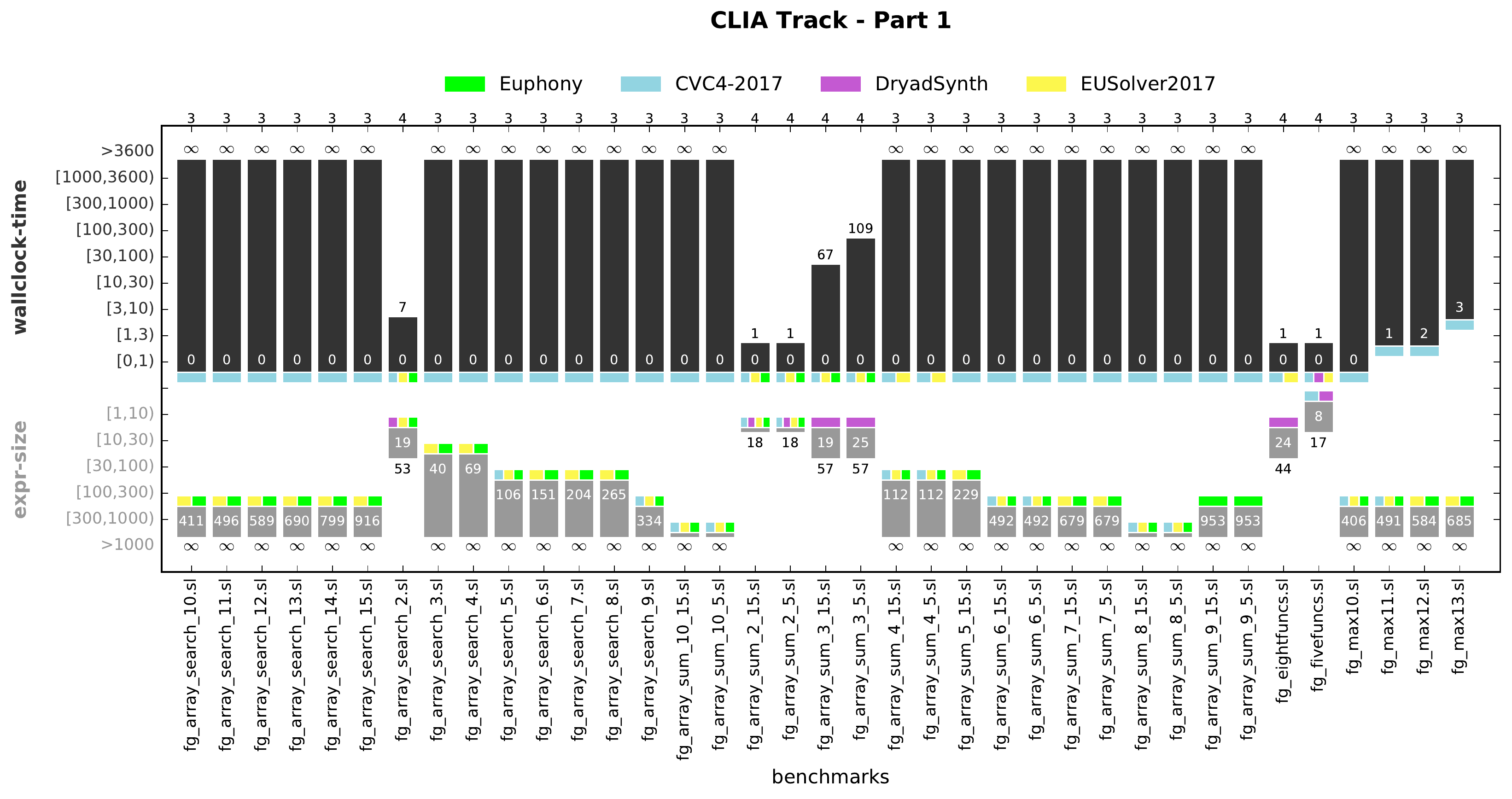} \\[3cm]
			\includegraphics[width=9.5in,bb=7 9 925 460]{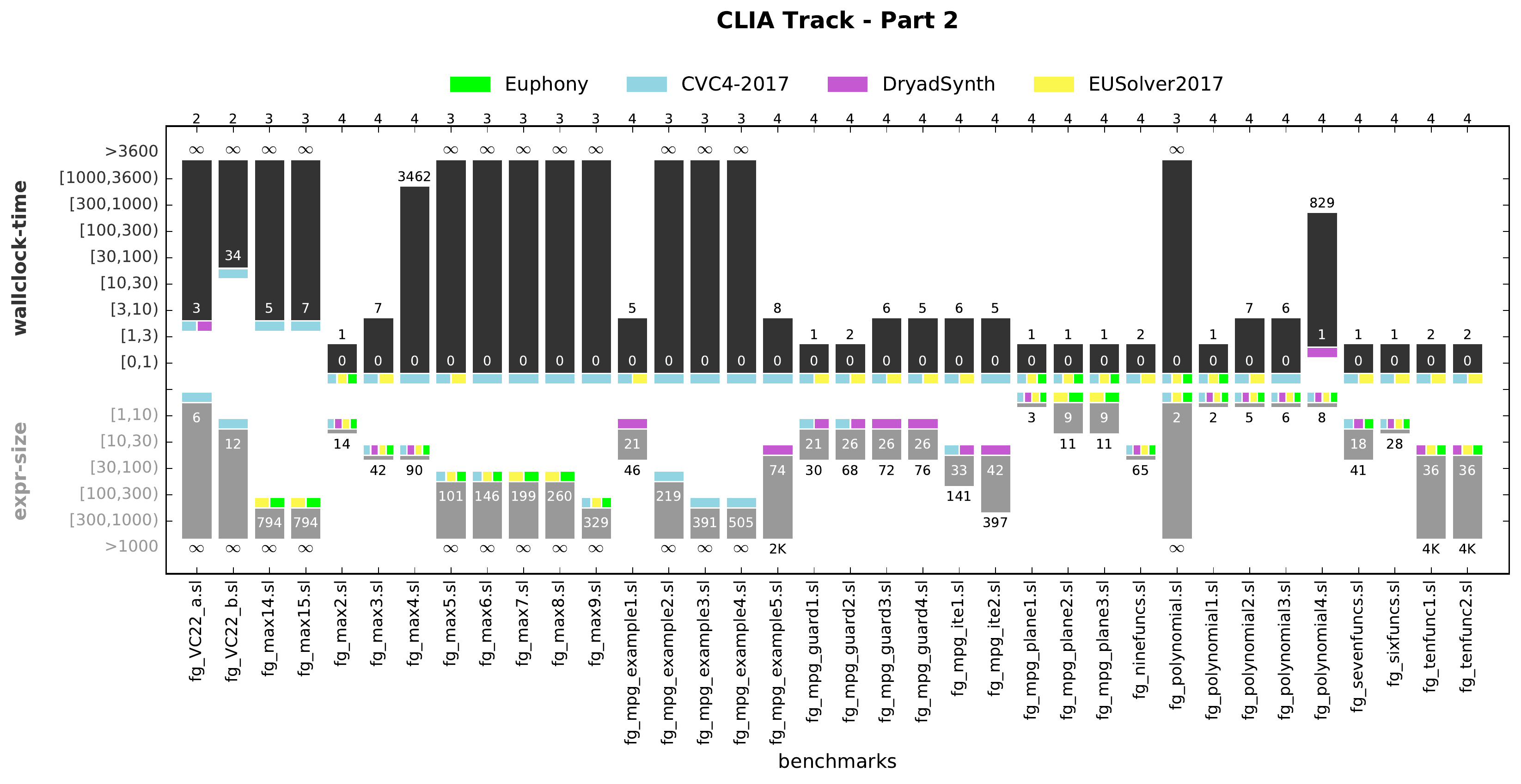} 
		\end{tabular}
	}}
	\caption{Evaluation of CLIA track benchmarks.}\label{fig:clia-results}
\end{figure*}

\begin{figure*}
\noindent\makebox[\textwidth]{
	\scalebox{0.6}{
		\begin{tabular}{c}
			\includegraphics[width=9.5in,bb=7 9 923 476]{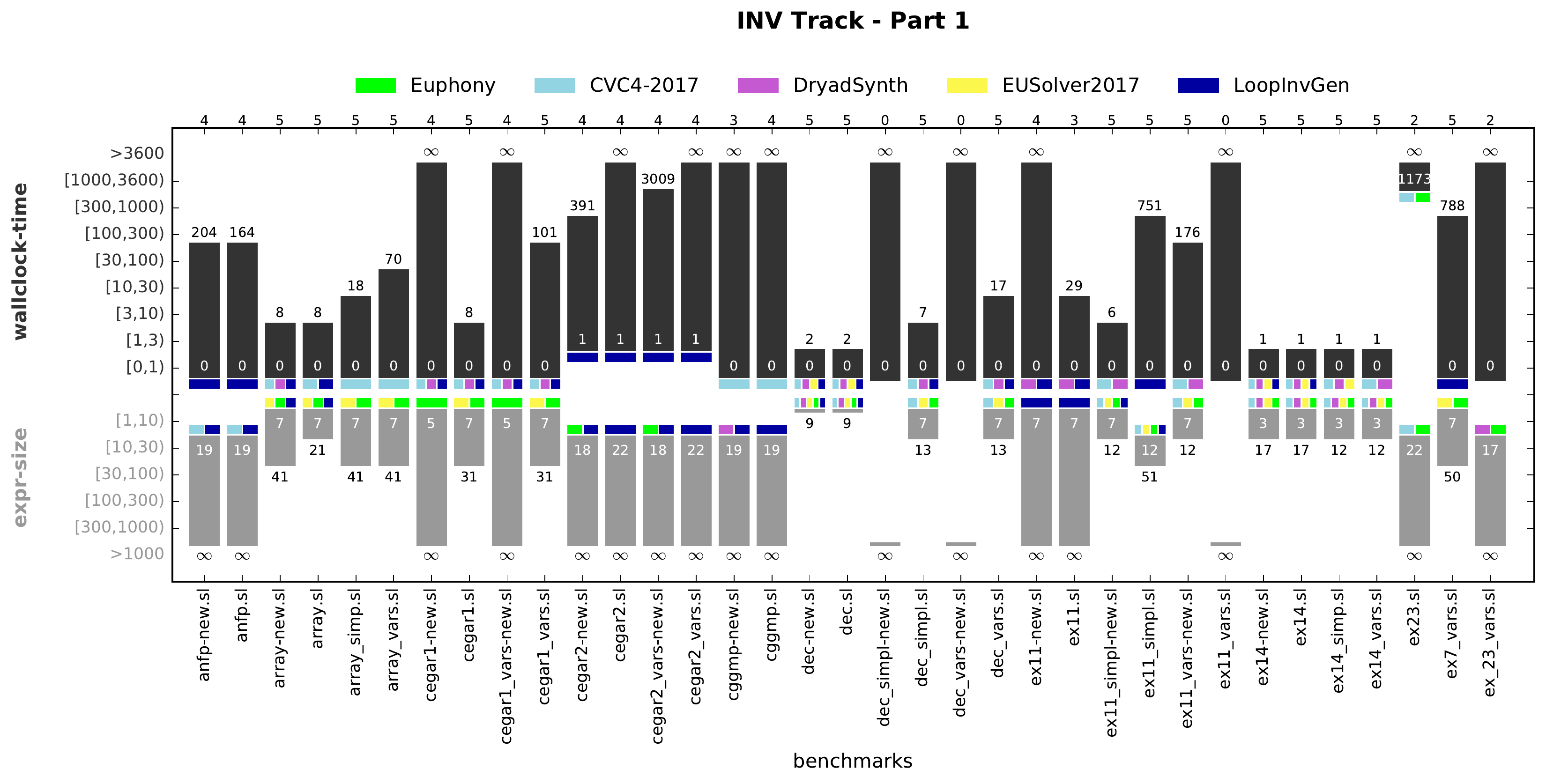} \\[3cm]
			\includegraphics[width=9.5in,bb=7 9 925 460]{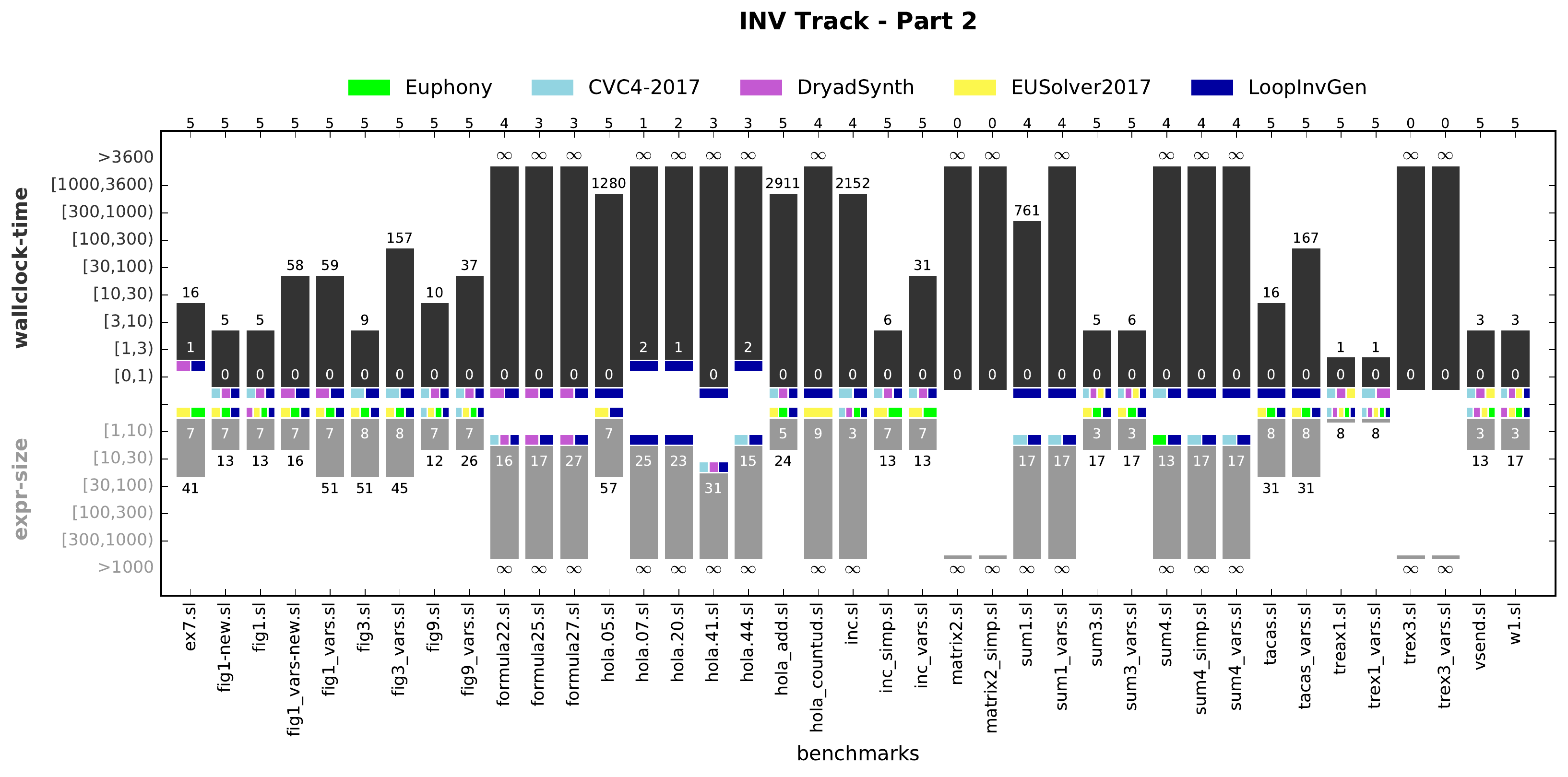} 
		\end{tabular}
	}}
	\caption{Evaluation of Invariant track benchmarks.}\label{fig:inv-results}
\end{figure*}

\begin{figure*}
\noindent\makebox[\textwidth]{
	\scalebox{0.6}{
		\begin{tabular}{c}
			\includegraphics[width=9.5in,bb=7 9 923 476]{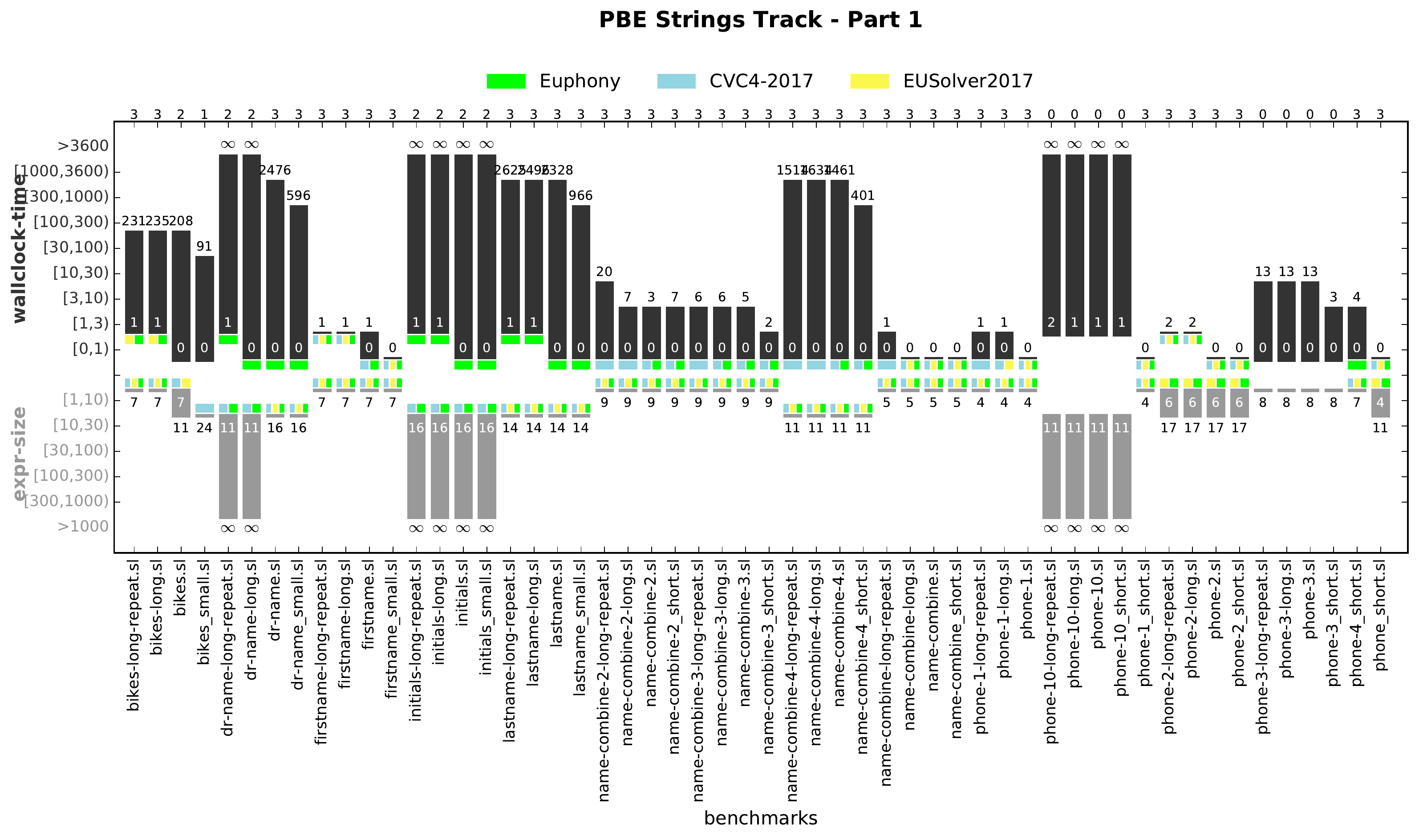} \\[3cm]
			\includegraphics[width=9.5in,bb=7 9 925 460]{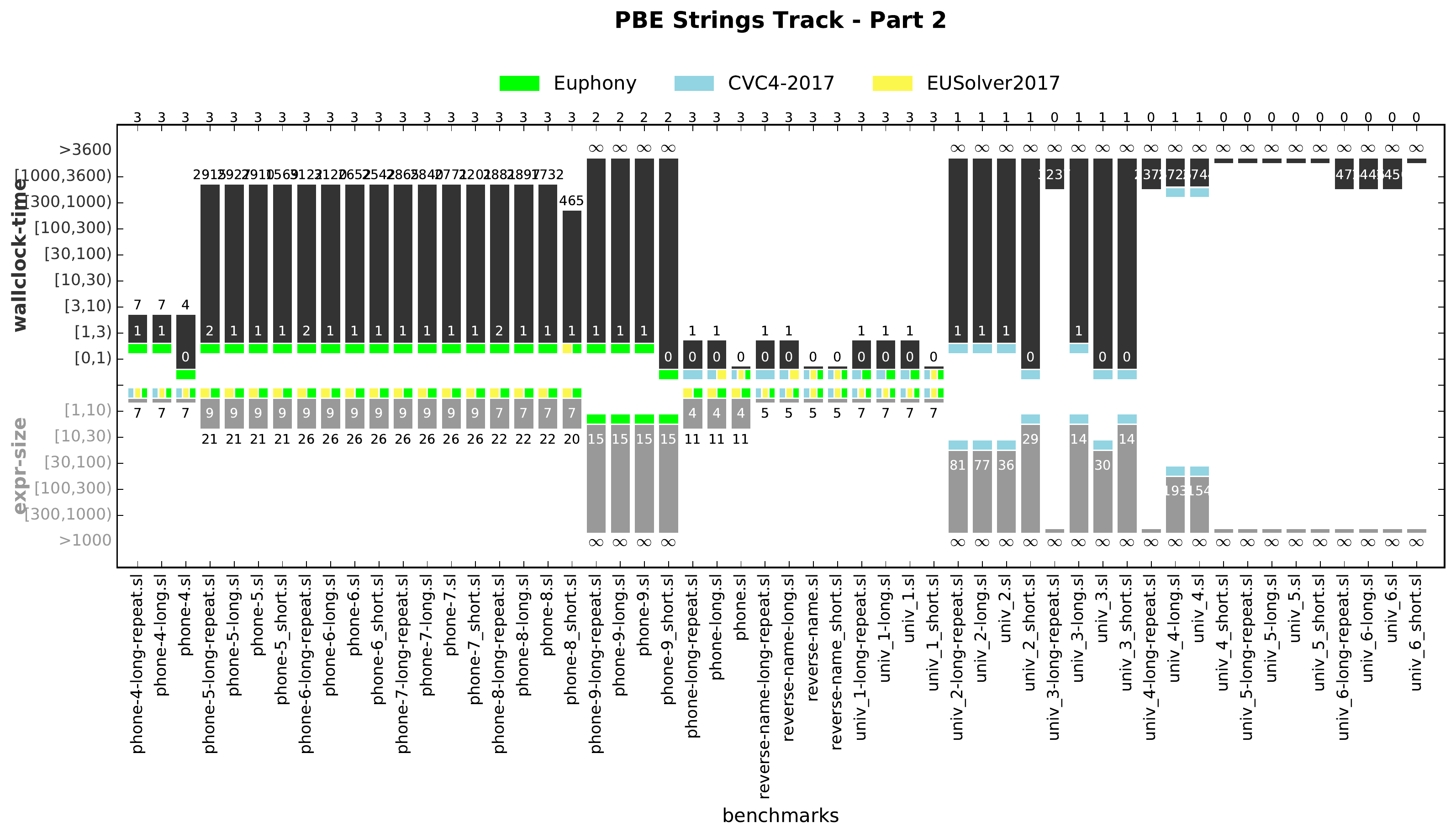} 
		\end{tabular}
	}}
	\caption{Evaluation of PBE Strings track benchmarks.}\label{fig:pbe-strings-results}
\end{figure*}

\section{Summary}
\label{sec:discussion}
This year's competition consisted of over 1500 benchmarks, 250 of which where contributed this year. Six solvers competed this year, out of which four by developers submitting a tool for SyGuS-Comp for the first time. All tools preformed remarkably, on both existing and new benchmarks. In particular, 65\% of the new benchmarks were solved.

An impressive progress was shown this year in solving the strings benchmarks of the programing by example track. Analyzing the features of benchmarks that are still hard to solve, we see that these include those with either (i) multiple functions to synthesize or (ii) where the specification invokes the functions with different parameters or (iii) those that use the \emph{let} expression for specifying auxiliary variables, or (iv) the grammar is very general consisting of much more operators than needed, or (v) the specification is partial in the sense that the domain of semantic solutions is not a singleton.

\bibliographystyle{eptcs}
\bibliography{bib-wdoi}

\begin{thebibliography}{10}
\providecommand{\bibitemdeclare}[2]{}
\providecommand{\surnamestart}{}
\providecommand{\surnameend}{}
\providecommand{\urlprefix}{Available at }
\providecommand{\url}[1]{\texttt{#1}}
\providecommand{\href}[2]{\texttt{#2}}
\providecommand{\urlalt}[2]{\href{#1}{#2}}
\providecommand{\doi}[1]{doi:\urlalt{http://dx.doi.org/#1}{#1}}
\providecommand{\bibinfo}[2]{#2}

\bibitemdeclare{inproceedings}{AlurBJMRSSSTU13}
\bibitem{AlurBJMRSSSTU13}
\bibinfo{author}{Rajeev \surnamestart Alur\surnameend},
  \bibinfo{author}{Rastislav \surnamestart Bod{\'{\i}}k\surnameend},
  \bibinfo{author}{Garvit \surnamestart Juniwal\surnameend},
  \bibinfo{author}{Milo M.~K. \surnamestart Martin\surnameend},
  \bibinfo{author}{Mukund \surnamestart Raghothaman\surnameend},
  \bibinfo{author}{Sanjit~A. \surnamestart Seshia\surnameend},
  \bibinfo{author}{Rishabh \surnamestart Singh\surnameend},
  \bibinfo{author}{Armando \surnamestart Solar{-}Lezama\surnameend},
  \bibinfo{author}{Emina \surnamestart Torlak\surnameend} \&
  \bibinfo{author}{Abhishek \surnamestart Udupa\surnameend}
  (\bibinfo{year}{2013}): \emph{\bibinfo{title}{Syntax-guided synthesis}}.
\newblock In: {\sl \bibinfo{booktitle}{Formal Methods in Computer-Aided Design,
  {FMCAD} 2013, Portland, OR, USA, October 20-23, 2013}}, pp.
  \bibinfo{pages}{1--8}.

\bibitemdeclare{inproceedings}{AlurCAV15}
\bibitem{AlurCAV15}
\bibinfo{author}{Rajeev \surnamestart Alur\surnameend}, \bibinfo{author}{Pavol
  \surnamestart Cern{\'{y}}\surnameend} \& \bibinfo{author}{Arjun \surnamestart
  Radhakrishna\surnameend} (\bibinfo{year}{2015}):
  \emph{\bibinfo{title}{Synthesis Through Unification}}.
\newblock In: {\sl \bibinfo{booktitle}{Computer Aided Verification - 27th
  International Conference, {CAV} 2015, San Francisco, CA, USA, July 18-24,
  2015, Proceedings, Part {II}}}, pp. \bibinfo{pages}{163--179},
  \doi{10.1007/978-3-319-21668-3_10}.

\bibitemdeclare{incollection}{SyGuSComp15}
\bibitem{SyGuSComp15}
\bibinfo{author}{Rajeev \surnamestart Alur\surnameend}, \bibinfo{author}{Dana
  \surnamestart Fisman\surnameend}, \bibinfo{author}{Rishabh \surnamestart
  Singh\surnameend} \& \bibinfo{author}{Armando \surnamestart
  Solar-Lezama\surnameend} (\bibinfo{year}{2015}):
  \emph{\bibinfo{title}{Results and Analysis of SyGuS-Comp'15}}.
\newblock In: {\sl \bibinfo{booktitle}{SYNT}}, \bibinfo{publisher}{EPTCS}, pp.
  \bibinfo{pages}{3--26}, \doi{10.4204/EPTCS.202.3}.

\bibitemdeclare{inproceedings}{AlurRU17}
\bibitem{AlurRU17}
\bibinfo{author}{Rajeev \surnamestart Alur\surnameend}, \bibinfo{author}{Arjun
  \surnamestart Radhakrishna\surnameend} \& \bibinfo{author}{Abhishek
  \surnamestart Udupa\surnameend} (\bibinfo{year}{2017}):
  \emph{\bibinfo{title}{Scaling Enumerative Program Synthesis via Divide and
  Conquer}}.
\newblock In: {\sl \bibinfo{booktitle}{Tools and Algorithms for the
  Construction and Analysis of Systems - 23rd International Conference, {TACAS}
  2017, Held as Part of the European Joint Conferences on Theory and Practice
  of Software, {ETAPS} 2017, Uppsala, Sweden, April 22-29, 2017, Proceedings,
  Part {I}}}, pp. \bibinfo{pages}{319--336},
  \doi{10.1007/978-3-662-54577-5_18}.

\bibitemdeclare{misc}{smtlib}
\bibitem{smtlib}
\bibinfo{author}{Clark \surnamestart Barrett\surnameend},
  \bibinfo{author}{Aaron \surnamestart Stump\surnameend} \&
  \bibinfo{author}{Cesare \surnamestart Tinelli\surnameend}:
  \emph{\bibinfo{title}{The SMT-LIB Standard – Version 2.0}}.

\bibitemdeclare{inproceedings}{EldibWW16}
\bibitem{EldibWW16}
\bibinfo{author}{Hassan \surnamestart Eldib\surnameend}, \bibinfo{author}{Meng
  \surnamestart Wu\surnameend} \& \bibinfo{author}{Chao \surnamestart
  Wang\surnameend} (\bibinfo{year}{2016}): \emph{\bibinfo{title}{Synthesis of
  Fault-Attack Countermeasures for Cryptographic Circuits}}.
\newblock In: {\sl \bibinfo{booktitle}{Computer Aided Verification - 28th
  International Conference, {CAV} 2016, Toronto, ON, Canada, July 17-23, 2016,
  Proceedings, Part {II}}}, pp. \bibinfo{pages}{343--363},
  \doi{10.1007/978-3-319-41540-6_19}.

\bibitemdeclare{inproceedings}{GNMR16}
\bibitem{GNMR16}
\bibinfo{author}{Pranav \surnamestart Garg\surnameend}, \bibinfo{author}{Daniel
  \surnamestart Neider\surnameend}, \bibinfo{author}{P.~\surnamestart
  Madhusudan\surnameend} \& \bibinfo{author}{Dan \surnamestart Roth\surnameend}
  (\bibinfo{year}{2016}): \emph{\bibinfo{title}{Learning invariants using
  decision trees and implication counterexamples}}.
\newblock In: {\sl \bibinfo{booktitle}{Proceedings of the 43rd Annual {ACM}
  {SIGPLAN-SIGACT} Symposium on Principles of Programming Languages, {POPL}
  2016, St. Petersburg, FL, USA, January 20 - 22, 2016}}, pp.
  \bibinfo{pages}{499--512}, \doi{10.1145/2837614.2837664}.

\bibitemdeclare{inproceedings}{repairbenchmarks}
\bibitem{repairbenchmarks}
\bibinfo{author}{Xuan{-}Bach~D. \surnamestart Le\surnameend},
  \bibinfo{author}{Duc{-}Hiep \surnamestart Chu\surnameend},
  \bibinfo{author}{David \surnamestart Lo\surnameend}, \bibinfo{author}{Claire
  \surnamestart {Le Goues}\surnameend} \& \bibinfo{author}{Willem \surnamestart
  Visser\surnameend} (\bibinfo{year}{2017}): \emph{\bibinfo{title}{{S3:}
  syntax- and semantic-guided repair synthesis via programming by examples}}.
\newblock In: {\sl \bibinfo{booktitle}{FSE}}, pp. \bibinfo{pages}{593--604},
  \doi{10.1145/3106237.3106309}.

\bibitemdeclare{misc}{ICEDT}
\bibitem{ICEDT}
\bibinfo{author}{Daniel \surnamestart Neider\surnameend},
  \bibinfo{author}{P.~\surnamestart Madhusudan\surnameend} \&
  \bibinfo{author}{Pranav \surnamestart Garg\surnameend}
  (\bibinfo{year}{2015}): \emph{\bibinfo{title}{ICE DT: Learning Invariants
  using Decision Trees and Implication Counterexamples}}.
\newblock \bibinfo{note}{Private Communication}.

\bibitemdeclare{article}{PadhiM17}
\bibitem{PadhiM17}
\bibinfo{author}{Saswat \surnamestart Padhi\surnameend} \&
  \bibinfo{author}{Todd~D. \surnamestart Millstein\surnameend}
  (\bibinfo{year}{2017}): \emph{\bibinfo{title}{Data-Driven Loop Invariant
  Inference with Automatic Feature Synthesis}}.
\newblock {\sl \bibinfo{journal}{CoRR}} \bibinfo{volume}{abs/1707.02029}.
\newblock \urlprefix\url{http://arxiv.org/abs/1707.02029}.

\bibitemdeclare{inproceedings}{PadhiSM16}
\bibitem{PadhiSM16}
\bibinfo{author}{Saswat \surnamestart Padhi\surnameend}, \bibinfo{author}{Rahul
  \surnamestart Sharma\surnameend} \& \bibinfo{author}{Todd~D. \surnamestart
  Millstein\surnameend} (\bibinfo{year}{2016}):
  \emph{\bibinfo{title}{Data-driven precondition inference with learned
  features}}.
\newblock In: {\sl \bibinfo{booktitle}{Proceedings of the 37th {ACM} {SIGPLAN}
  Conference on Programming Language Design and Implementation, {PLDI} 2016,
  Santa Barbara, CA, USA, June 13-17, 2016}}, pp. \bibinfo{pages}{42--56},
  \doi{10.1145/2908080.2908099}.

\bibitemdeclare{article}{RaghothamanU14}
\bibitem{RaghothamanU14}
\bibinfo{author}{Mukund \surnamestart Raghothaman\surnameend} \&
  \bibinfo{author}{Abhishek \surnamestart Udupa\surnameend}
  (\bibinfo{year}{2014}): \emph{\bibinfo{title}{Language to Specify
  Syntax-Guided Synthesis Problems}}.
\newblock {\sl \bibinfo{journal}{CoRR}} \bibinfo{volume}{abs/1405.5590}.

\bibitemdeclare{inproceedings}{ReynoldsDKTB15}
\bibitem{ReynoldsDKTB15}
\bibinfo{author}{Andrew \surnamestart Reynolds\surnameend},
  \bibinfo{author}{Morgan \surnamestart Deters\surnameend},
  \bibinfo{author}{Viktor \surnamestart Kuncak\surnameend},
  \bibinfo{author}{Cesare \surnamestart Tinelli\surnameend} \&
  \bibinfo{author}{Clark~W. \surnamestart Barrett\surnameend}
  (\bibinfo{year}{2015}): \emph{\bibinfo{title}{Counterexample-Guided
  Quantifier Instantiation for Synthesis in {SMT}}}.
\newblock In: {\sl \bibinfo{booktitle}{Computer Aided Verification - 27th
  International Conference, {CAV} 2015, San Francisco, CA, USA, July 18-24,
  2015, Proceedings, Part {II}}}, pp. \bibinfo{pages}{198--216},
  \doi{10.1007/978-3-319-21668-3_12}.

\bibitemdeclare{misc}{ReynoldsT17}
\bibitem{ReynoldsT17}
\bibinfo{author}{Andrew \surnamestart Reynolds\surnameend} \&
  \bibinfo{author}{Cesare \surnamestart Tinelli\surnameend}
  (\bibinfo{year}{2017}): \emph{\bibinfo{title}{{SyGuS} Techniques in the Core
  of an {SMT} Solver}}.
\newblock \bibinfo{note}{To appear in this issue.}

\bibitemdeclare{inproceedings}{starexec}
\bibitem{starexec}
\bibinfo{author}{Aaron \surnamestart Stump\surnameend}, \bibinfo{author}{Geoff
  \surnamestart Sutcliffe\surnameend} \& \bibinfo{author}{Cesare \surnamestart
  Tinelli\surnameend} (\bibinfo{year}{2014}): \emph{\bibinfo{title}{StarExec:
  {A} Cross-Community Infrastructure for Logic Solving}}.
\newblock In: {\sl \bibinfo{booktitle}{Automated Reasoning - 7th International
  Joint Conference, {IJCAR} 2014, Held as Part of the Vienna Summer of Logic,
  {VSL} 2014, Vienna, Austria, July 19-22, 2014. Proceedings}}, pp.
  \bibinfo{pages}{367--373}, \doi{10.1007/978-3-319-08587-6_28}.

\bibitemdeclare{inproceedings}{UdupaRDMMA13}
\bibitem{UdupaRDMMA13}
\bibinfo{author}{Abhishek \surnamestart Udupa\surnameend},
  \bibinfo{author}{Arun \surnamestart Raghavan\surnameend},
  \bibinfo{author}{Jyotirmoy~V. \surnamestart Deshmukh\surnameend},
  \bibinfo{author}{Sela \surnamestart Mador{-}Haim\surnameend},
  \bibinfo{author}{Milo M.~K. \surnamestart Martin\surnameend} \&
  \bibinfo{author}{Rajeev \surnamestart Alur\surnameend}
  (\bibinfo{year}{2013}): \emph{\bibinfo{title}{{TRANSIT:} specifying protocols
  with concolic snippets}}.
\newblock In: {\sl \bibinfo{booktitle}{{ACM} {SIGPLAN} Conference on
  Programming Language Design and Implementation, {PLDI} '13, Seattle, WA, USA,
  June 16-19, 2013}}, pp. \bibinfo{pages}{287--296},
  \doi{10.1145/2462156.2462174}.

\end{thebibliography}
\end{document}